\documentclass[12pt]{article} 
\usepackage{epsfig,ldd_art,latexsym}
\input{epsf.tex}
\newcommand{\One}{1\kern-4.5pt1}
\newcommand{\lapprox}{\raisebox{-0.5ex}{$\ 
\stackrel{\textstyle<}{\textstyle\sim}\ $}}
\newcommand{\gapprox}{\raisebox{-0.5ex}{$\ 
\stackrel{\textstyle>}{\textstyle\sim}\ $}}
\begin{document}

\addtolength{\baselineskip}{0.20\baselineskip}

\rightline{ SWAT/01/313}

\hfill September 2001
\vspace{24pt}

\centerline{\bf NUMERICAL PORTRAIT OF A RELATIVISTIC THIN FILM BCS SUPERFLUID}


\vspace{18pt}

\centerline{\bf Simon Hands$^a$, Biagio Lucini$^b$ and Susan
Morrison$^a$}

\vspace{15pt}

\centerline{\sl $^a$Department of Physics, University of Wales Swansea,}
\centerline{\sl Singleton Park, Swansea SA2 8PP, U.K.}

\centerline{\sl $^b$Department of Theoretical Physics, 1 Keble Road, Oxford OX1
3NP, U.K.}

\vfill


\centerline{{\bf Abstract}}

\noindent
{\narrower 
We present results of numerical simulations of the 2+1$d$ 
Nambu -- Jona-Lasinio model with a non-zero
baryon chemical potential $\mu$ including the effects of a diquark source term.
Diquark condensates, susceptibilities and masses are measured as functions
of source strength $j$. The results suggest that diquark condensation does
not take place in the high density phase $\mu>\mu_c$, 
but rather that the condensate
scales non-analytically with $j$ implying a line of critical points
and long range phase coherence. Analogies are
drawn with the low temperature phase of the $2d$ $XY$ model.
The spectrum of the spin-${1\over2}$ sector is also studied yielding
the quasiparticle dispersion relation. There is no evidence for a non-zero gap;
rather the results are characteristic of a normal Fermi liquid with Fermi
velocity less than that of light. We conclude that the high density phase of the
model describes a relativistic gapless thin film BCS superfluid.
}



\vfill

\newpage

\section{Introduction}

Spontaneous symmetry breaking in particle physics was pre-dated by the 
Bardeen-Cooper-Schrieffer (BCS) mechanism for superconductivity
in metals at low temperature \cite{BCS}, which predicts 
a ground state in which a macroscopic fraction 
of the electrons in the vicinity of the Fermi surface
reside in spin-0 bound states known as Cooper pairs. In field theoretic terms
\cite{Gorkov} the Cooper pairs form 
a condensate which alter the symmetry of the ground state;
in the case of superconductivity U(1) electromagnetic
gauge invariance 
is spontaneously
broken, leading to the Meissner effect, ie.
exclusion of magnetic field from a superconducting
sample due to surface screening currents.
The ideas of BCS have been incorporated into particle physics in two distinct
directions. Firstly, particle -- anti-particle pair condensation 
$\langle\bar\psi\psi\rangle\not=0$ was suggested
as a means of breaking the global chiral symmetries responsible for keeping
fermion masses small; Goldstone's theorem then predicts light 
weakly-interacting bosonic states which can be identified with pions, whose
masses are considerably less than the nucleon, the lightest strongly-interacting
fermion. The
resulting model provides a reasonable description of low-energy
strong-interaction phenomena \cite{NJL}. Secondly, condensation of an 
elementary Higgs field has of course 
been invoked as a mechanism for electroweak
gauge symmetry breaking, imbuing gauge bosons (as well as
fermionic matter fields)
with non-zero mass in precise analogy with the Meissner effect.

In recent years the BCS mechanism has returned to particle physics in a new
guise in the context of the fundamental theory of the strong interaction,
QCD, at high density. For baryon charge
densities $n_B\sim O(1)\mbox{fm}^{-3}$, it is believed that 
chiral symmetry is restored and
nucleons dissociate into quarks.
The resulting ground state is thought to
be {\sl quark matter\/} in which to first approximation the dominant degrees of
freedom are relativistic degenerate quarks forming a Fermi sphere with 
Fermi momentum $k_F\approx350-400$MeV. Such conditions are conceivably found in
the cores of neutron stars. However, since the force between quarks due to eg.
one-gluon exchange is attractive, this simple picture is unstable with respect
to a BCS scenario in which condensation of {\sl diquark pairs\/} occurs
\cite{diquarks}. Since the $qq$ wavefunction is gauge non-singlet, the resulting
ground state renders some or all of the gluons massive, an effect known as 
``color superconductivity''. The resulting dynamically
generated mass scale or ``gap'' $\Delta$ is predicted to be 
$O(100)$MeV \cite{Berges}, 
and hence comparable with the consituent quark scale.

Unfortunately, theoretical studies of color superconductivity are to date
limited to perturbative and self-consistent methods \cite{diquarks,Berges,
Pis}; there is no
systematic method of performing non-perturbative QCD calculations in the high
density regime because of the notorious ``sign problem'', ie. the measure of
the Euclidean path integral becomes complex once baryon chemical potential 
$\mu\not=0$, making the importance sampling techniques traditionally used
in numerical simulations of lattice gauge theory ineffective. There are,
however, strongly-interacting model field theories where this difficulty can be
circumvented. One is QCD with just two colors, in which $qq$ baryons and 
$q\bar q$ mesons fall into multiplets
related by enhanced global symmetries. Some of the multiplets contain Goldstone
bosons, so that the methods of chiral perturbation theory can be applied
\cite{KSTVZ}.
Baryonic matter $n_B>0$ forms for chemical potential $\mu\gapprox m_\pi/2$
\cite{KSTVZ,istvan,KSHM}.
The resulting ground state is a superfluid
of light but strongly bound $qq$ states which form a Bose-Einstein condensate.
There is no Fermi surface in this regime, and the BCS description is
inappropriate. 

Another possibility, to be studied in the present paper, 
are four-fermi models such as
the Nambu -- Jona-Lasinio (NJL) model \cite{NJL}, 
in which it can be shown that the effects of adding
``conjugate quarks'' $q^c$ 
to make the path integral real and positive have little impact on
the physics: at low density light 
Goldstone states
arising as a result of chiral symmetry breaking can only form in
mesonic $q\bar q$ channels, whereas baryonic $qq^c$ bound states
remain massive, ie. at the constituent quark scale \cite{Barbour}. This means
that unlike in physical three-color QCD with conjugate quarks, it is possible 
for simulations
of four-fermi models to maintain a separation of scales between $m_\pi$ and
the critical $\mu_c$ at which chiral symmetry is restored and baryonic matter
induced into the ground state \cite{kim}. In this case the model 
does not reproduce any of the physics of confinement, but has a
Fermi surface for $n_B>0$.

Because the $qq$ interaction is attractive, diquark condensation is expected 
in the high density phase of each of the models described above.
In both cases, however, the relevant $\langle qq\rangle\not=0$
is gauge singlet, 
meaning that the ground state is not superconducting, but rather {\sl
superfluid\/}. In field theoretic terms, a superfluid forming by BCS
condensation is characterised by 
a ground state which does not respect a global symmetry of the underlying
action, in this case the U(1) corresponding to baryon number, which is  thus
no longer a good quantum number. Fermionic excitations above the
Fermi surface are 
a superposition of particle and hole states, and require energy
$\geq\Delta$ to excite.
Finally, because a continuous
global symmetry is spontaneously broken, Goldstone's theorem applies and
massless 
diquark states are expected in the excitation spectrum. Physically these
result both in a long-ranged interaction between the 
vortex excitations found in a rotating superfluid, and in propagating 
waves of temperature variation known as {\sl second sound\/} \cite{TT,AM}.

The two known superfluids are liquid $^4$He at kelvin and liquid $^3$He
at milli-kelvin temperatures. $^4$He is a boson and is
naturally treated using a complex scalar field theory, superfluidity 
arising via a Bose-Einstein condensation.
Note, however, that 
a fundamental description would treat $^4$He as a tightly-bound state of 
fermionic constituents, not too dissimilar in spirit to Two-Color QCD.
$^3$He by contrast is a fermion, and superfluidity in this case is believed to
arise via a BCS instability resulting in a condensation of weakly-bound
Cooper pairs.
We might thus consider superfluidity in the NJL model as a relativistic
generalisation of this phenomenon. It is important to note, though, that due to
short distance repulsion between helium atoms the BCS wavefunction in $^3$He is
actually $p$-wave, resulting in ground states described by a complicated order
parameter and many interesting topological excitations \cite{Volovik}.
In the NJL model studied here, the corresponding $qq$ wavefunction is a scalar
$s$-wave, and any superfluid might be expected to behave more like $^4$He.

Superfluidity in a relativistic model similar 
to the NJL model has been studied using mean
field techniques in \cite{Pis}. In our previous work \cite{HM,HLM}
we have attempted to identify 
superfluidity in the 2+1$d$ NJL model using non-perturbative
numerical lattice simulations. Apart from the obvious computational gain,
we chose this particular dimensionality 
because the model has a non-trivial continuum
limit \cite{RWP,HKK}. Therefore, in contrast to effective descriptions 
such as the Landau-Ginzburg theory, the condensed matter described in this
approach is formed from the elementary 
quanta of an interacting field theory.
Our results have not supported the expected scenario outlined above, 
raising the question of whether important physics is neglected in the
self-consistent approach.
Although there is evidence for enhanced diquark pairing in the
scalar isosinglet channel in \cite{HM}, we have not succeeded in finding
an unambiguous signal for a condensate $\langle qq\rangle\not=0$.
Rather, in the high density phase the condensate appears to vanish
non-analytically as a function of diquark source strength $j$, 
suggesting critical behaviour
\cite{HLM}. Studies of the excitation spectrum in the spin-${1\over2}$ sector
reveal a sharp Fermi surface and no evidence for a gap $\Delta\not=0$
\cite{HLM}. The purpose of the present paper is to present these results in 
greater depth, and to attempt to interpret them. As we shall see, it may be 
possible to attribute the unconventional signals to the specifically
two-dimensional nature of the system being studied, which thus bears many
resemblances to superfluidity observed in thin helium films \cite{Nelson}. 
We will argue that
neither long range order $\langle qq\rangle\not=0$ nor a gap $\Delta\not=0$ are
necessary attributes of a superfluid. Instead, the critical behaviour observed
in \cite{HLM} results from long range {\sl coherence\/} 
in the phase of the diquark
wavefunction, which appears to be a sufficient condition for thin film
superfluidity \cite{KT}.

In Sec.~\ref{sec:model} we review the formulation and numerical simulation 
of the lattice NJL model in 2+1$d$ with non-zero chemical potential $\mu$,
paying particular attention to the introduction of diquark source terms $jqq$
via the
use of a {\sl Gor'kov basis\/} \cite{Gorkov}. It is thus possible to define
diquark observables which are
measurable on a finite system; Sec.~\ref{sec:diquarks} reviews numerical results
for the diquark condensate $\langle qq(j)\rangle$, 
the associated susceptibilites, and diquark masses. Critical behaviour 
in the high density phase $n_B>0$ is 
identified after extrapolating results for the first two quantities to the
zero-temperature (ie. $L_t\to\infty$) limit, leading to consistent estimates for
the critical exponent conventionally denoted $\delta$ which vary with $\mu$. 
In Sec.~\ref{sec:crit}
this behaviour is discussed in analogy with that of the $2d$ $XY$ model, in 
which long range order is washed out by spin wave excitations and which
also displays critical behaviour in a continuous parameter range. 
It is argued that in such circumstances 
persistent flow, the defining property of a superfluid, can only be disrupted
by excitations costing infinite energy in the thermodynamic limit.
Sec.~\ref{sec:quasi} presents the results of a study of the dispersion
relation $E(k)$ of spin-${1\over2}$ {\sl quasiparticle\/} excitations in the
dense phase, revealing a sharp Fermi surface for the first time using lattice
methods. There is no evidence for particle-hole mixing or a non-vanishing
gap as the source strength $j\to0$. Instead, the results in this sector
are consistent with a normal Fermi liquid of the type first discussed by Landau
\cite{Landau2,Landau}; 
in particular it is possible to estimate both Fermi momentum
$k_F$ and velocity $\beta_F$ as functions of $\mu$ and to show that these depart
from their free-field values, yielding information on quasiparticle
interactions. Conclusions and suggestions for further work are outlined in
Sec.~\ref{sec:conc}

\section{The Lattice Model}
\label{sec:model}

\subsection{Formulation and Symmetries}
The model studied in this paper is a lattice transcription of the NJL
model in 2+1 dimensions, identical to that studied in
\cite{HM,HLM}. It is defined by the Euclidean action
\begin{eqnarray}
S=S_{fer}+S_{bos}&:&\nonumber\\
S_{fer}=\sum_x\bar\chi M[\Phi]\chi 
+ j\chi^{tr}\tau_2\chi + \bar\jmath\bar\chi
\tau_2\bar\chi^{tr}&;&
S_{bos}={1\over g^2}\sum_{\tilde x}\mbox{tr}\Phi^\dagger\Phi,
\label{eq:action}
\end{eqnarray}
where $\chi$, $\bar\chi$ are isospinor fermionic 
fields defined on the 
sites $x$ of a 2+1$d$ lattice, and $\Phi\equiv\sigma+i\vec\pi.\vec\tau$
is a $2\times2$ matrix of bosonic auxiliary
fields living on the dual sites $\tilde x$.
The kinetic operator $M$ has 
the standard form for staggered lattice fermions interacting with
scalar fields \cite{HKK}:
\begin{eqnarray}
M_{xy}^{pq}[\Phi] & = & {1\over2}\delta^{pq}
\left[(\mbox{e}^\mu\delta_{yx+\hat0}-\mbox{e}^{-\mu}\delta_{yx-\hat0})
+\sum_{\nu=1,2}\eta_\nu(x)(\delta_{yx+\hat\nu}-\delta_{yx-\hat\nu})
+2m\delta_{xy}\right]\nonumber\\
& + &
{1\over8}\delta_{xy}
\sum_{\langle\tilde x,x\rangle}\biggl(\sigma(\tilde x)\delta^{pq}
+i\varepsilon(x)\vec\pi(\tilde x).\vec\tau^{\,pq}\biggr).
\label{eq:M}
\end{eqnarray}
The parameters are bare fermion mass $m$, baryon chemical potential
$\mu$, and coupling $g^2$. The $\vec\tau$ are Pauli matrices normalised
to $\mbox{tr}(\tau_i\tau_j)=2\delta_{ij}$
acting on isopin indices $p,q=1,2$.
The symbols $\eta_\nu(x)$ denote the phases $(-1)^{x_0+\cdots+x_{\nu-1}}$,
$\varepsilon(x)$ the phase $(-1)^{x_0+x_1+x_2}$,
and $\langle\tilde x,x\rangle$ the set of 8
dual sites neighbouring $x$. Integration over the auxiliary $\Phi$
fields
leads to an equivalent action in terms of fermions
which self-interact via a four-point contact term proportional to $g^2$,
corresponding to the interaction of the NJL model
\cite{HK}.

In addition to the usual NJL interactions, Eqn. (\ref{eq:action}) contains
diquark and anti-diquark terms proportional to source strengths $j$ and
$\bar\jmath$ respectively. These have been introduced to enable the measurement
of the diquark condensate $\langle \chi^{tr}\tau_2\chi\rangle$ on a finite
system, 
in precise analogy to the role of the bare mass $m$ in the measurement of the
chiral condensate $\langle\bar\chi\chi\rangle$. To proceed, we define the
bispinor $\Psi^{tr}=(\bar\chi^{tr},\chi)$, and rewrite the fermion action as a
quadratic form $S_{fer}=\Psi^{tr}{\cal
A}\Psi$, where in this Gor'kov basis 
the antisymmetric matrix ${\cal A}$ is 
\begin{equation}
{\cal A}=\left(\matrix{\bar\jmath\tau_2&{1\over2}M\cr
         -{1\over2}M^{tr}&j\tau_2\cr}\right).
\end{equation}
The fermion fields may then be integrated out to yield the following Euclidean
path integral:
\begin{equation}
Z=\int D\sigma D\vec\pi\; \mbox{Pf}(2{\cal A}[\Phi,j,\bar\jmath])
\exp-S_{bos}[\Phi],
\label{eq:Z}
\end{equation}
where the pfaffian $\mbox{Pf}(Q)\equiv\sqrt{\mbox{det}Q}$. Note that this
expression differs from the (incorrect) 
version given in \cite{HLM} by a physically irrelevant factor of two; 
for convenience we will stick
with the current notation, but note that if the source is interpreted as 
a Majorana mass $\lambda$, then $j=\lambda/2$ \cite{KSHM}.

The model described by the action (\ref{eq:action},\ref{eq:M}) has an
$\mbox{SU}(2)_L\otimes\mbox{SU}(2)_R\otimes\mbox{U}(1)_B$ global symmetry.
Defining projection operators ${\cal P}_{e/o}={1\over2}(1\pm\varepsilon(x))$ 
onto even and odd sublattices respectively, we have
\begin{equation}
\chi\mapsto({\cal P}_eU+{\cal P}_oV)\chi;\;\;\;
\bar\chi\mapsto\bar\chi({\cal P}_eV^\dagger+{\cal P}_oU^\dagger);\;\;\;
\Phi\mapsto V\Phi U^\dagger\;\;[U,V\in\mbox{SU(2)}]:
\label{eq:su2xsu2}
\end{equation}
\begin{equation}
\chi\mapsto e^{i\alpha}\chi;\;\;\;\bar\chi\mapsto\bar\chi e^{-i\alpha}
\;\;[e^{i\alpha}\in\mbox{U(1)}_B].
\label{eq:u1}
\end{equation}
The symmetry (\ref{eq:su2xsu2}) is broken to a diagonal SU(2)$_V$ of isospin
with $U\equiv V$ in (\ref{eq:su2xsu2}), 
either explicitly by a bare fermion mass $m\not=0$, or spontaneously
by the generation of a chiral condensate $\langle\bar\chi\chi\rangle\not=0$ by
the model's dynamics. For $\mu=0$ this occurs for a sufficiently strong
coupling $g^2>g_c^2\simeq1.0a$, where $a$ is the physical lattice spacing 
\cite{HKK}. In the chirally broken phase the fermions have a dynamically
generated mass $\Sigma\simeq\langle\sigma\rangle\equiv
{g^2\over2}\langle\bar\chi\chi\rangle$. Since $\Sigma a\to0$ 
as $g^2\to g_c^2$, 
a continuum limit may be taken at this critical point. A remarkable feature
of the 2+1$d$ NJL model is that the continuum theory so obtained 
remains interacting \cite{RWP,HKK}. As in our previous studies \cite{HM,HLM}, 
the simulations
in this paper were performed with $g^2=2.0$ corresponding to $\Sigma
a=0.71$, implying that we are rather far from the continuum limit.

\begin{figure}[htb]
\begin{center}
\epsfig{file=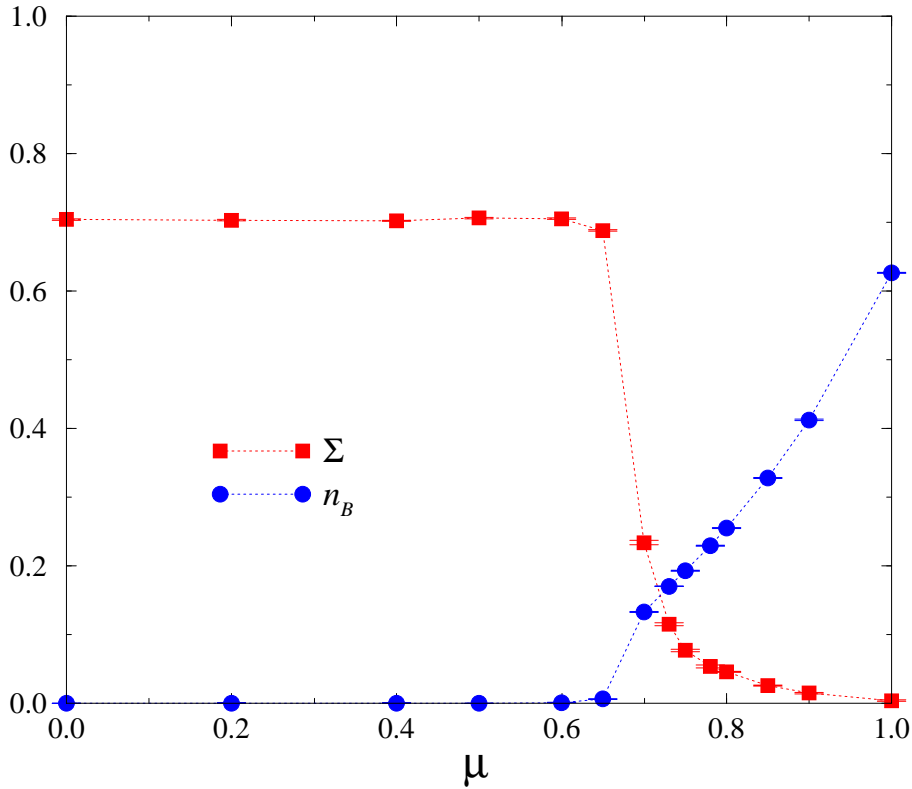, width =12cm}
\end{center}
\caption{
\label{fig:condensates}
Chiral condensate $\Sigma$ and baryon charge density $n_B$ as a function of
$\mu$ for a $16^2\times24$ system with $g^2=2.0$, $m=0.01$, $j=0$.}
\end{figure}
For $\mu\not=0$ the model is known to exhibit a strong first-order transition to
a chirally symmetric phase \cite{RWP2}; for our realisation this occurs
at a critical $\mu_c\simeq\Sigma\simeq0.65$, as shown in
Fig~\ref{fig:condensates} \cite{HM}. Chiral symmetry 
restoration is accompanied at this point by the onset of a
non-vanishing density of baryon charge in the ground state, signalled by 
a condensate
\begin{equation}
n_B={1\over 2V}{{\partial\ln Z}\over\partial\mu}
={1\over4}\langle\bar\chi(x)e^\mu\chi(x+\hat0)
+\bar\chi(x)e^{-\mu}\chi(x-\hat0)\rangle>0.
\end{equation}
Existing numerical evidence suggests these two {\it a priori\/}
distinct transitions are coincident
\cite{Costas}.
For $\mu>\mu_c$, $\Sigma\simeq0$, and the density follows the approximate
form $n_B\propto\mu^2$, the 
behaviour expected for massless states populating 
a 2 dimensional Fermi sphere of radius 
$E_F=\mu$, until it gets close to its saturation value of one quark of each
isospin per lattice site.
For our realisation, $n_B\simeq0.25$ quarks of each isospin label per site 
at $\mu=0.8$, 
corresponding to a physical density of about 80 fm$^{-2}$ assuming a constituent
quark mass $\Sigma$ of 300 MeV; by $\mu=0.9$ this has risen to $n_B\simeq0.4$,
corresponding to 125 fm$^{-2}$.

The question which 
will occupy us in this paper is the nature of the high density phase 
present for $\mu>\mu_c$, and in
particular whether the U(1)$_B$ symmetry (\ref{eq:u1}) is spontaneously 
broken by the generation of a diquark condensate which we will 
generically denote
by $\langle qq\rangle\not=0$. In a BCS condensation, the 
participating diquark pairs come from the neighbourhood of the Fermi surface.
The resulting ground state is separated from excited states
by an energy gap $\Delta$ analogous to the constituent quark mass $\Sigma$ in a
chirally-broken vacuum; mean-field calculations of a phenomenologically
inspired 3+1$d$ NJL-type model predict that for quark matter
$\Delta$ is of the same
order as $\Sigma$ \cite{Berges}. 
As well as being massive, the quasiparticle excitations 
above the ground state carry indefinite baryon charge due to the U(1)$_B$
breaking. Physically, this means that the quasiparticle is a coherent
superposition of particle and hole states. In Ref. \cite{HM} diquark timeslice
correlators
$\sum_{\vec x}\langle qq(\vec0,0)\bar q\bar q(\vec x,t)\rangle$ 
in various plausible condensation 
channels were studied for $\mu>\mu_c$, and
evidence for pairing found, in the form of a plateau whose height did not
decrease with Euclidean time separation $t$, for the scalar SU(2)$_L\otimes
$SU(2)$_R$ singlet
channel $qq=\chi^{tr}\tau_2\chi$. However, the naive interpretation of diquark
condensation via the clustering hypothesis, namely that
\begin{equation}
\lim_{t\to\infty}\langle qq(0)\bar q\bar q(t)\rangle=\vert\langle
qq\rangle\vert^2,
\end{equation}
was ruled out because the plateau height did not scale in the expected way, ie.
extensively in the spatial volume $L_s^2$. To clarify the situation, Ref.
\cite{HLM} introduced diquark
source terms, as in (\ref{eq:action}), 
making direct measurements of $\langle
qq\rangle$ possible; 
we now review the `standard' signals which might be expected 
if diquark condensation occurs.

Firstly we define diquark operators $qq_\pm$ via
\begin{equation}
qq_\pm(x)={1\over2}\left[\chi^{tr}(x)\tau_2\chi(x)\pm
\bar\chi(x)\tau_2\bar\chi^{tr}(x)\right],
\label{eq:qq}
\end{equation}
with corresponding source strengths $j_\pm=j\pm\bar\jmath$. It is readily
checked that $qq_\pm$ are invariant under SU(2)$_L\otimes$SU(2)$_R$ rotations
(\ref{eq:su2xsu2}), but rotate into each other under U(1)$_B$
(\ref{eq:u1}). In terms of 4-component spinors $\psi$, 
the operators (\ref{eq:qq})
may be written \cite{HM}
\begin{equation}
qq_\pm=-i\left[\psi^{tr}({\cal C}\gamma_5\otimes\tau_2\otimes\tau_2)\psi\pm
\bar\psi({\cal C}\gamma_5\otimes\tau_2\otimes\tau_2)\bar\psi^{tr}\right],
\end{equation}
where the first matrix in the tensor product acts on spinor indices,
the second on a 2-component flavor structure which is implicit in the staggered
fermion approach \cite{BB}, and the third on the explicit isospin index 
introduced in
(\ref{eq:action}). The charge conjugation
matrix ${\cal C}$ is defined by ${\cal C}\gamma_\mu{\cal
C}^{-1}=-\gamma_\mu^*$. The diquark condensate is now given by
\begin{equation}
\langle qq_+\rangle={1\over V}{{\partial\ln Z}\over\partial j_+}=
{1\over4V}\langle\mbox{tr}\tau_2{\cal A}^{-1}\rangle, 
\label{eq:qq+}
\end{equation}
and is calculable using standard lattice techniques, such as the use of a
stochastic estimator for the diagonal elements of the inverse matrix. The
non-vanishing of (\ref{eq:qq+}) in the limit $j_+\to0$ is a criterion for the
spontaneous breakdown of U(1)$_B$ symmetry.
Furthermore, if we
define susceptibilities 
\begin{equation}
\chi_\pm=\sum_x\langle qq_\pm(0)qq_\pm(x)\rangle, 
\label{eq:susc}
\end{equation}
then it is straightforward to derive a Ward identity analogous to the axial Ward
identity for the pion propagator:
\begin{equation}
\chi_-\vert_{j_-=0}={{\langle qq_+\rangle}\over j_+}.
\label{eq:ward}
\end{equation}
On the assumption that the dominant contribution to $\chi_-$ is from a simple
pole, then $qq_-$ couples to a Goldstone mode whose mass $M_-$
vanishes in the zero source limit as $\surd j_+$. 
If we similarly attribute $\chi_+$ to a ``Higgs mode'',
then the ratio $\chi_+/\chi_-$ provides an alternative means of distinguishing
possible symmetry-breaking scenarios in the limit $j_+\to0$:
\begin{equation}
R=\lim_{j_+\to0}-{\chi_+\over\chi_-}=
\cases{1,& if U(1)$_B$ manifest;\cr
                                         0,& if U(1)$_B$ broken.\cr}
\label{eq:ratio}
\end{equation}
In Sec.~\ref{sec:diquarks} we will present numerical results for these
quantities and discuss to what extent the above considerations 
help in describing the high density phase of the 2+1$d$ NJL model. 

\subsection{The Simulation}

Numerical simulation of the path integral (\ref{eq:Z}) requires some discussion
of how to deal with the pfaffian. First let us find the condition
that $\mbox{det}2{\cal A}$ is real. Using the property of a block square matrix
\begin{equation}
{\rm det}\left(\matrix{X&Y\cr W&Z\cr}\right)=
{\rm det}X{\rm det}(Z-WX^{-1}Y),
\end{equation}
and the property $\tau_2M\tau_2=M^*$ which follows from (\ref{eq:M}), we deduce
\begin{equation}
\mbox{det}2{\cal A}=\mbox{det}(4j\bar\jmath+M^\dagger M), 
\label{eq:pfaffian}
\end{equation}
and is hence real
and positive if $j\bar\jmath$ is chosen real and positive. In all our work we
choose $j=\bar\jmath$ real which satisfies this condition. It follows that
$\mbox{Pf}(2{\cal A})$ is real. In fact, one can go further and argue that it 
is also positive as follows. In the limit $j,\bar\jmath\to0$, $\mbox{Pf}(2{\cal
A})$ reduces to $\mbox{det}M$, which can be proven both real and positive  
using an argument, identical to that used for SU(2) 
lattice gauge theory with staggered 
quarks in the fundamental representation \cite{istvan}, showing that any 
complex eigenvalue of $M$ is accompanied in the spectrum by its conjugate, 
and any purely real eigenvalue is doubly degenerate. Relation
(\ref{eq:pfaffian}), however, shows that $\mbox{det}2{\cal A}$ can only increase
and hence cannot change sign once $j\bar\jmath>0$; it follows that
$\mbox{Pf}(2{\cal A})$ can be consistently chosen real and positive \cite{KSHM}.

Despite this reassuring property, in our simulation we chose to use
$\mbox{det}^{1\over2}({\cal A}^\dagger{\cal A})\propto\mbox{Pf}^2(2{\cal A})$ 
as the measure, corresponding to 
two staggered lattice fermion species, 
for consistency with the model of \cite{HM} which
is recovered in the limit $j\to0$. It can be shown that 
in the continuum limit the model contains $N_f=4$ species of four-component 
Dirac fermions \cite{BB}. The simulation is performed using a hybrid molecular
dynamics ``R'' algorithm \cite{ray}, in which the square root is taken by 
inserting a factor of ${1\over2}$ in the force term derived from a local action;
note that we were able to debug and tune the code by also
checking against an exact algorithm for the case $N_f=8$. In all cases we used
a molecular dynamics timestep $\Delta\tau=0.04$, and never saw any evidence
of departure from equipartition of energy.
We performed
simulations on lattice sizes $L_s^2\times L_t=16^3$, 
$16^2\times24$, $24^3$, $32^3$, and in one
case $48^3$,
with the coupling $g^2$ fixed to 2.0 as described above, and bare Dirac
mass $m$
fixed to 0.01 to assist with the identification of chirally broken and restored
phases. A typical run is over $O(400)$ HMD time units with mean refreshment
interval $1.0$; data were taken every two units. The cost
of the simulation rises considerably
in the chirally-restored phase where the 
diagonal elements of $M$, proportional to $\langle\sigma\rangle$, are small,
particularly as $j\to0$. The $48^3$ point at $\mu=0.8, j=0.025$ required 
approximately 40 SGI Origin2000 processor days.

In Sec.~\ref{subs:pq} we review the behaviour of the model as a function
of $\mu$, and present results for $\langle qq\rangle$ taken in the ``partially
quenched'' approximation in which $j\not=0$ only in the
measurement, and the simulation performed using an exact algorithm with
$j=0$. Our studies with full pfaffian dynamics, presented in
Secs.~\ref{subs:dqc},\ref{subs:susc} and \ref{subs:diqm}, focussed on two
representative points in the chirally symmetric low density phase at
$\mu=0.0,0.2$, and on two values in the high density chirally symmetric phase
$\mu=0.8,0.9$. We used $j=\bar\jmath$ 
ranging in value from 0.3 down to
0.025. In our studies of the quasiparticle spectrum discussed in 
Sec.~\ref{sec:quasi} we performed runs on $32^3$
at 4 additional values of $\mu\in(0.8,0.9)$.

\section{Numerical Results for Diquark Observables}
\label{sec:diquarks}

\subsection{Partially Quenched Results}
\label{subs:pq}

\begin{figure}[htb]
\begin{center}
\epsfig{file=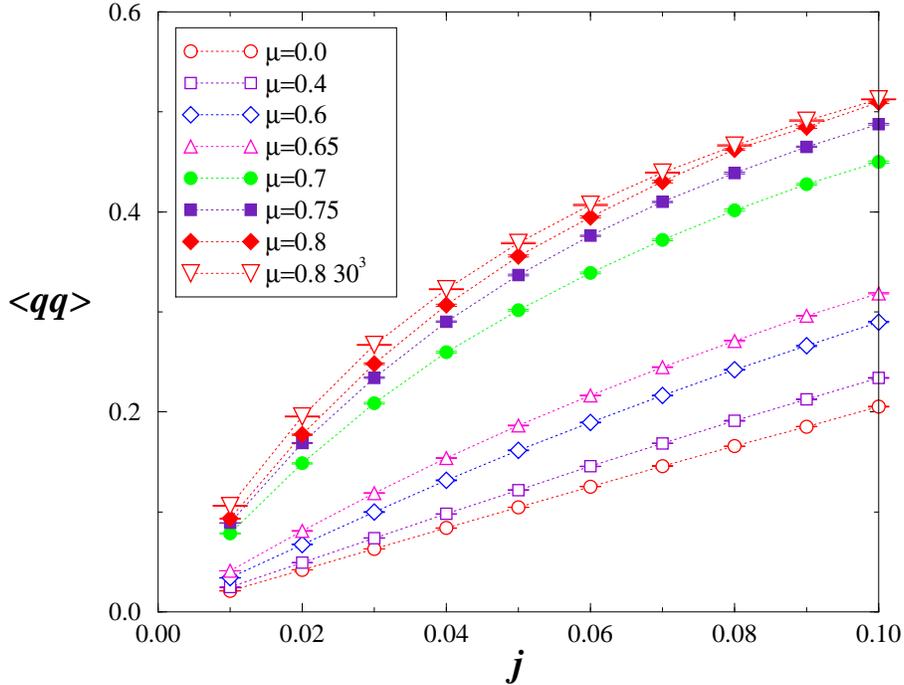, width =12cm}
\end{center}
\caption{
\label{fig:qqpc}
Partially quenched results for the 
diquark condensate $\langle\chi^{tr}\tau_2\chi\rangle$ as 
a function of source strength $j$ for various values of $\mu$; unless otherwise
shown, data was taken on a
$16^2\times24$ system.}
\end{figure}
In this section we discuss the direct numerical measurement of diquark 
condensates $\langle qq\rangle$, which the Gor'kov basis discussed in the
previous section 
makes possible.
To warm up we consider the partially quenched approximation, in which 
the source strength $j$ is set to zero in the updating of the $\{\Phi\}$
configuration, which thus proceeds via an exact Hybrid Monte Carlo algorithm as
in \cite{HM}, 
but is non-zero in the measurement routine so that $\langle qq\rangle$
can be measured via (\ref{eq:qq+}). Since at a given $\mu$ a single simulation 
serves for all $j$, this approach is fairly cheap and hence good
coverage of the $\mu$ axis is practicable. Our results are shown in
Fig.~\ref{fig:qqpc}, where open symbols denote data taken in the low density
phase $\mu<\mu_c$, and filled symbols are from the dense phase. For $\mu<\mu_c$
$\langle qq_+\rangle$ varies approximately linearly with $j$, implying that a 
smooth extrapolation to the origin is possible, 
and hence the condensate vanishes in
the $j\to0$ limit. A striking jump occurs between $\mu=0.65$ and $\mu=0.7$, 
and for values of $\mu$ in the dense phase $\langle qq(j)\rangle$ is markedly
more curved. This behaviour cannot be taken as evidence
of diquark condensation, however; one should expect discontinuities in all 
physical observables on different sides of a first order phase transition. 
Despite the
curvature of the lines a smooth extrapolation to the origin consistent 
with unbroken baryon number symmetry at high density is still plausible. 
Another possibility, suggested by the $\mu=0.8$ data from a $30^3$ 
lattice shown in Fig.~\ref{fig:qqpc}, 
is that the symmetry breaking is masked by a 
finite-volume suppression as $j\to0$.
To explore the behaviour for $\mu>\mu_c$ in more depth data from lattices
with several distinct $L_s$ and $L_t$ are needed. We chose to perform this using
``unitary'' data generated using the correct measure (\ref{eq:Z}) for a 
limited number of different $\mu$, as described next. 

\subsection{Diquark Condensate}
\label{subs:dqc}

\begin{figure}[p]
\begin{center}
\epsfig{file=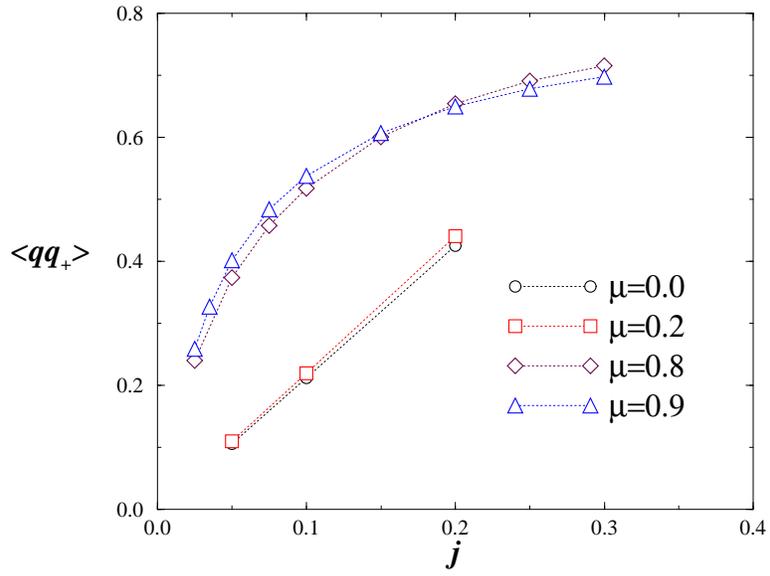, width =10cm}
\end{center}
\caption{
\label{fig:qq_unit}
Results for 
$\langle qq_+\rangle$ vs. 
$j$ for various values of $\mu$ from 
a full simulation on a $32^3$
system.}
\end{figure}
\begin{figure}[p]
\begin{center}
\epsfig{file=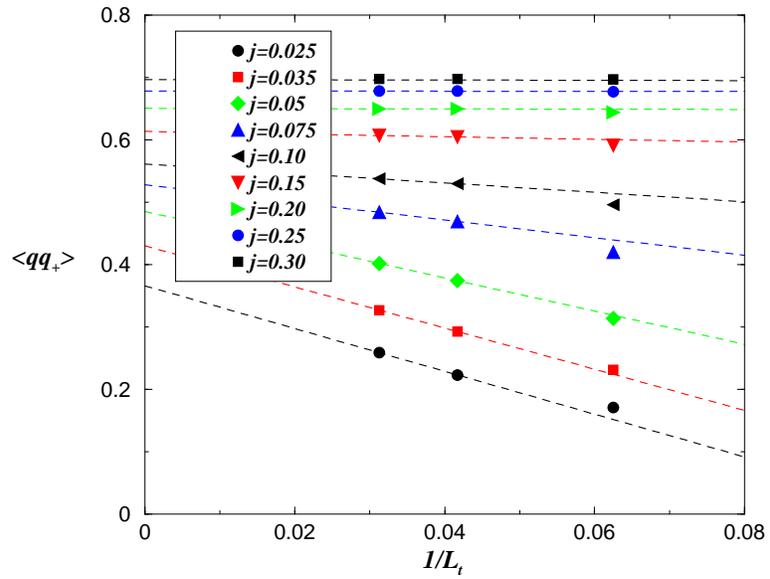, width =10cm}
\end{center}
\caption{
\label{fig:qq_extrap}
Extrapolation of 
$\langle qq_+\rangle$ at $\mu=0.9$ to the thermodynamic limit.}
\vskip 2 truecm
\end{figure}
\begin{table}[htb]
\caption{Values of $\langle qq_+(j)\rangle$ on various lattice sizes at
$\mu=0.8$.}
\label{tab:Ltscaling}
\smallskip
\begin{tabular*}{\textwidth}{@{}l@{\extracolsep{\fill}}llllll}
\hline
$j$ &  $16^3$ & $16^2\times24$ & $24^3$ & $32^3$ &  $48^3$ \\
\hline
0.025 & 0.1713(3) & 0.2158(3) &  0.2140(2) & 0.2400(2) &  0.2619(1)\\
0.035 & 0.2328(3) & $-$      &  0.2801(3) & $-$  & $-$\\
0.05  & 0.3125(4) & 0.3581(3) &  0.3572(3) & 0.3737(2) & $-$ \\
0.075 & 0.4165(5) & 0.4497(3) &  0.4492(3) & 0.4576(2) & $-$ \\
0.1   & 0.4921(5) & 0.5137(3) &  0.5133(4) & 0.5176(2) & $-$ \\
0.15  & 0.5910(6) & 0.5993(3) &  0.6002(3) & 0.6003(2) & $-$ \\
0.2   & 0.6506(5) & 0.6547(3) &  0.6542(3) & 0.6544(2) & $-$ \\
0.25  & 0.6896(5) & 0.6905(3) &  0.6910(3) & 0.6906(2) & $-$ \\
0.3   & 0.7148(5) & 0.7158(3) &  0.7160(3) & 0.7155(2) & $-$ \\
\hline
\end{tabular*}
\end{table}

In Fig.~\ref{fig:qq_unit} we show $\langle qq_+\rangle$ data taken from
simulations using the full pfaffian measure (\ref{eq:Z}) at two 
values of $\mu$ from each phase on a $32^3$ lattice.
The results resemble those of the partially quenched approach shown in
Fig.~\ref{fig:qqpc} both qualitatively and quantitatively. The
curvature of $\langle qq_+(j)\rangle$ in the dense phase seems 
to become more pronounced with increased $\mu$, to the extent that by
$j\simeq0.2$ the results
at $\mu=0.9$ actually undershoot those at $\mu=0.8$. As remarked in the previous
section, there are significant finite volume effects in this phase.
Fig.~\ref{fig:qq_extrap} shows $\mu=0.9$ data 
from simulations on $16^3$, $24^3$ and
$32^3$ lattices. The equivalent data for $\mu=0.8$, including a single point
from $48^3$, is tabulated in
Table~\ref{tab:Ltscaling} and plotted as  Fig.~2 of \cite{HLM}.
Empirically, we find by comparing data from
$16^3$, $16^2\times24$ and $24^3$ lattices that 
the dominant correction on a $L_s^2\times L_t$ system appears
due to finite $L_t$,
suggesting a specifically thermal origin. This motivates an extrapolation
to the thermodynamic limit which is linear
in $1/L_t$;
at smaller $j$, however, the data depart significantly from this trend.

\begin{figure}[htb]
\begin{center}
\epsfig{file=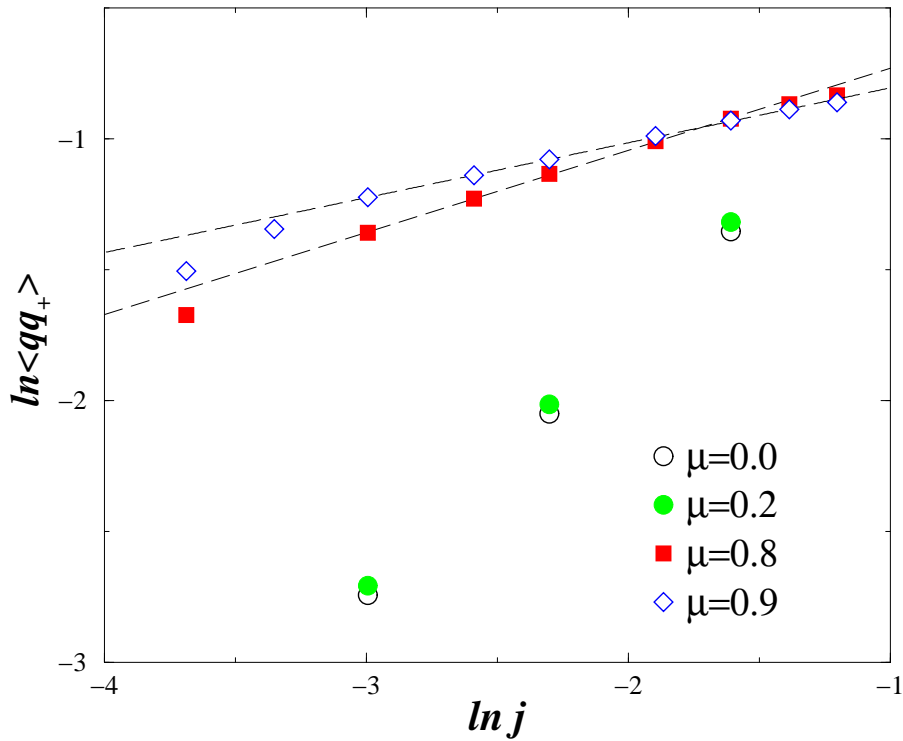, width =12cm}
\end{center}
\caption{
\label{fig:lnln}
$\ln\langle qq_+\rangle$ vs. $\ln j$, showing evidence for power-law scaling in
the dense phase.}
\end{figure}

Assuming a $1/L_t$ scaling, we extrapolated the data from $24^3$ and $32^3$
lattices to
estimate $\langle qq_+(j)\rangle$ in the thermodynamic limit. The results
are shown on a log-log plot in Fig.~\ref{fig:lnln}, together with unextrapolated
data from the chirally broken phase at $\mu=0.0,0.2$. Remarkably, there is a 
reasonably wide interval $j\in[0.05,0.2]$ within which the plot is approximately
linear, indicative of a power-law scaling
\begin{equation}
\langle qq_+(j)\rangle\propto j^\alpha.
\label{eq:powerlaw}
\end{equation}
Fits to (\ref{eq:powerlaw}) in this range yield $\alpha=0.314(2)$ for $\mu=0.8$,
with $\chi^2/\mbox{dof}=2.3$, and $\alpha=0.213(3)$ for $\mu=0.9$, with 
$\chi^2/\mbox{dof}=0.4$. In both cases this is clearly distinct 
from the linear (ie. $\alpha=1$) behaviour observed at low density.
For $j$ outside the fitted range, the data 
start to fall below the fitted line; we ascribe this to scaling violations
for $j\gapprox0.25$, as perhaps revealed by the crossing of the curves 
in Fig.~\ref{fig:qq_unit}, and for $j\lapprox0.035$ 
to non-thermal finite volume effects, eg. due to an insufficiently large
explicit Majorana mass, as perhaps indicated by the different
systematics of the $16^3$ point in Fig.~\ref{fig:qq_extrap}. Unfortunately
our resources have not permitted further systematic study of this point.

Assuming the validity of the form (\ref{eq:powerlaw}), we draw two conclusions.
Firstly, the non-analytic behaviour is reminiscent of the power-law scaling
observed at a critical point of a thermodynamic system. For a spin system 
at its critical temperature the spontaneous magnetisation ${\cal M}$ scales with
applied magnetic field $h$ as ${\cal M}\propto h^{1\over\delta}$ \cite{Ma}; 
for a
fermionic model exhibiting chiral symmetry breaking the equivalent 
relation is $\langle\bar\psi\psi\rangle\propto m^{1\over\delta}$ \cite{HKK}.
We thus identify critical scaling, with $\delta\equiv\alpha^{-1}$. Secondly, 
Fig.~\ref{fig:qqpc} leads us to expect that 
critical behaviour is generic in the dense phase, but with the exponent
$\delta$ varying continuously with chemical potential $\mu$, 
taking the value $\delta\approx3$ at $\mu=0.8$ and $\approx5$ at $\mu=0.9$.
This suggests a line of critical points for $\mu>\mu_c$. The origins of such 
behaviour and its consequences for superfluidity will be elaborated in 
Sec.~\ref{sec:crit}. In an attempt to find further evidence for
criticality, however, we now switch attention from one- to
two-point functions in a study of the various susceptibilities.

\subsection{Susceptibilities}
\label{subs:susc}

\begin{figure}[p]
\begin{center}
\epsfig{file=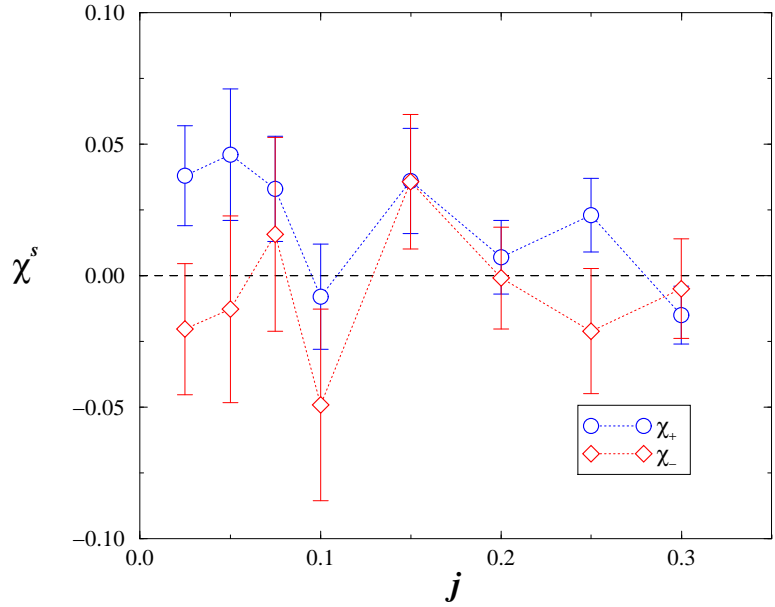, width =10cm}
\end{center}
\caption{
\label{fig:chi_sing}
$\chi^s$ vs. $j$ for $\mu=0.8$ on a $32^3$ lattice.}
\end{figure}

Next we examine the diquark susceptibilities $\chi_\pm$ defined by
(\ref{eq:susc}), which may be expanded to
\begin{eqnarray}
\chi_\pm={1\over4}\sum_x\langle\chi^{tr}\tau_2\chi(0)\chi^{tr}\tau_2\chi(x)
&+&\bar\chi\tau_2\bar\chi^{tr}(0)\bar\chi\tau_2\bar\chi^{tr}(x)\rangle\nonumber
\\
\pm\langle\chi^{tr}\tau_2\chi(0)\bar\chi\tau_2\bar\chi^{tr}(x)&+&
\bar\chi\tau_2\bar\chi^{tr}(0)\chi^{tr}\tau_2\chi(x)\rangle.
\label{eq:chil}
\end{eqnarray}
A  generic susceptibility
may be expressed as the sum of two connected contributions
corresponding to the two possible
Wick contractions,
\begin{equation}
\chi=\left[\langle(\mbox{tr}\Gamma{\cal G}_{xx})^2\rangle-
\langle\mbox{tr}\Gamma{\cal
G}_{xx}\rangle^2\right]+\langle\mbox{tr}
{\cal G}_{0x}\Gamma{\cal G}^{tr}_{0x}\Gamma\rangle\equiv
\chi^s+\chi^{ns},
\end{equation}
where  ${\cal G}={\cal A}^{-1}$ is the Gor'kov propagator and 
$\Gamma$ projects out the appropriate components. 
By analogy with meson physics we label these contributions ``singlet''
and ``non-singlet'' respectively. Estimates for
$\chi^s_\pm$ are made with
the same stochastic method used for $\langle qq_+\rangle$, and are plotted
for $\mu=0.8$ on a $32^3$ lattice in Fig.~\ref{fig:chi_sing}. Apart from the
observation that $\chi^s_-\approx\chi^s_+$, no other trend is 
apparent in the data, which are noisy and quite possibly consistent
with zero. In the following we ignore $\chi^s_\pm$ and assume
$\chi_\pm\simeq\chi^{ns}_\pm$. 
This is in marked contrast with the behaviour observed
in studies of chiral symmetry breaking in 2+1$d$ fermionic models, where 
singlet contributions to the relevant susceptibility are significant 
\cite{Thirring} or even dominant
\cite{Barbour}.
\begin{figure}[p]
\begin{center}
\epsfig{file=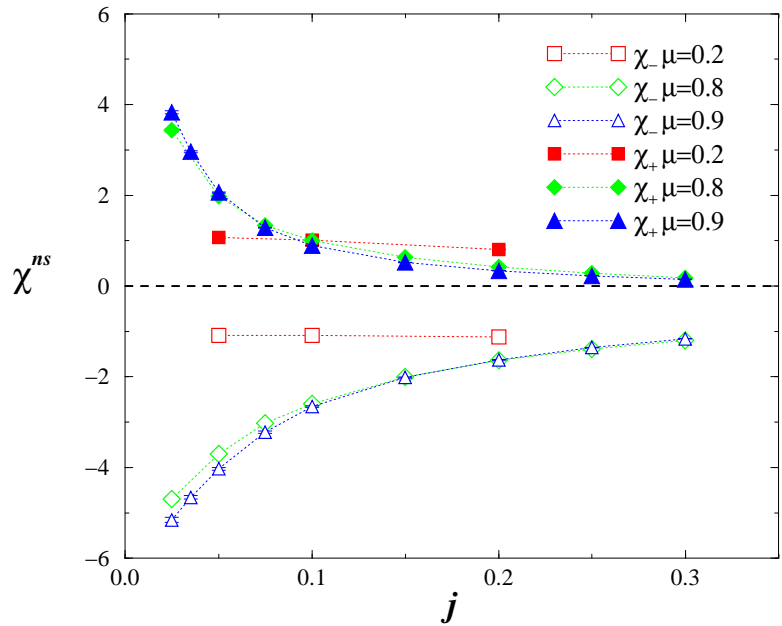, width =10cm}
\end{center}
\caption{
\label{fig:chi_nonsing}
$\chi^{ns}_+$ (filled) and $\chi^{ns}_-$ (open) 
vs. $j$ for various $\mu$ on a $32^3$ lattice.}
\end{figure}

Restricting attention to the non-singlet pieces, 
it is not hard to show using the properties of 
${\cal G}$ reviewed in Sec.~\ref{sec:quasi} below, that the first
expectation value on the right hand side of (\ref{eq:chil}) 
is negative and vanishes in 
the limit $j\to0$, whereas the second is positive and in fact 
corresponds to the diquark correlator
examined for $j=0$ in \cite{HM}. We conclude that
$\vert\chi_-\vert>\vert\chi_+\vert$. Data for $\chi^{ns}_\pm$
from a $32^3$ lattice are shown in Fig.~\ref{fig:chi_nonsing}.
Data at $\mu=0.0$ are
indistinguishable from those at $\mu=0.2$ on the scale plotted.
Note the difference of scale on the vertical axis between
Figs.~\ref{fig:chi_sing} and \ref{fig:chi_nonsing}.  We should also comment that
the $\chi^{ns}_-$ data when checked against $\langle qq_+\rangle$ saturate the
Ward identity (\ref{eq:ward}) within errors. Both observations justify our
neglect of $\chi^s_\pm$.

\begin{figure}[htb]
\begin{center}
\epsfig{file=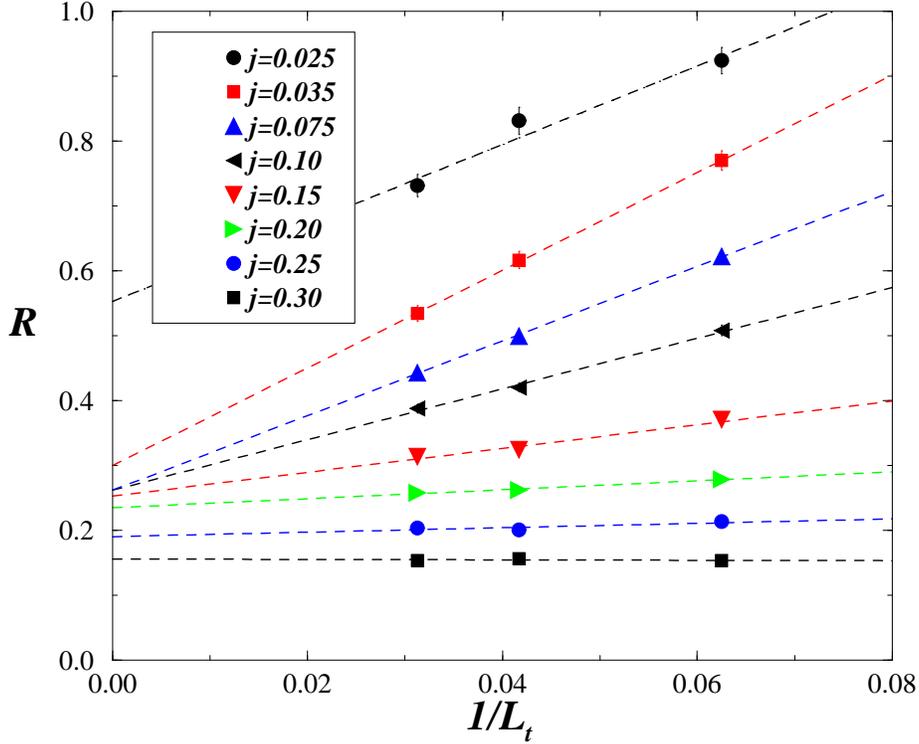, width =12cm}
\end{center}
\caption{
\label{fig:R_extrap}
Susceptibility ratio $R$ at $\mu=0.8$ for various lattice sizes.}
\end{figure}

In a conventional symmetry breaking
scenario $\chi_-$ should diverge in the thermodynamic and $j\to0$ limits
according to (\ref{eq:ward}), whereas $\chi_+$ could in
principle remain finite. 
The ratio $R$ defined in 
(\ref{eq:ratio}) is expected to vanish as $j\to0$
if U(1)$_B$ symmetry is spontaneously
broken by the ground state, and to approach unity 
if the symmetry remains manifest. In order to investigate this it is once
again necessary to take account of finite volume effects. There is no
appreciable effect for $\mu<\mu_c$, but in the dense
phase the variation with lattice size is considerable, as shown in
Fig.~\ref{fig:R_extrap}.
Once again, an extrapolation $\propto L_t^{-1}$ 
seems plausible, and indeed in this case a linear fit to the data from all three
available volumes proved acceptable. The accumulation of the resulting
intercepts in the region $R\simeq0.3$ is striking. 

\begin{figure}[htb]
\begin{center}
\epsfig{file=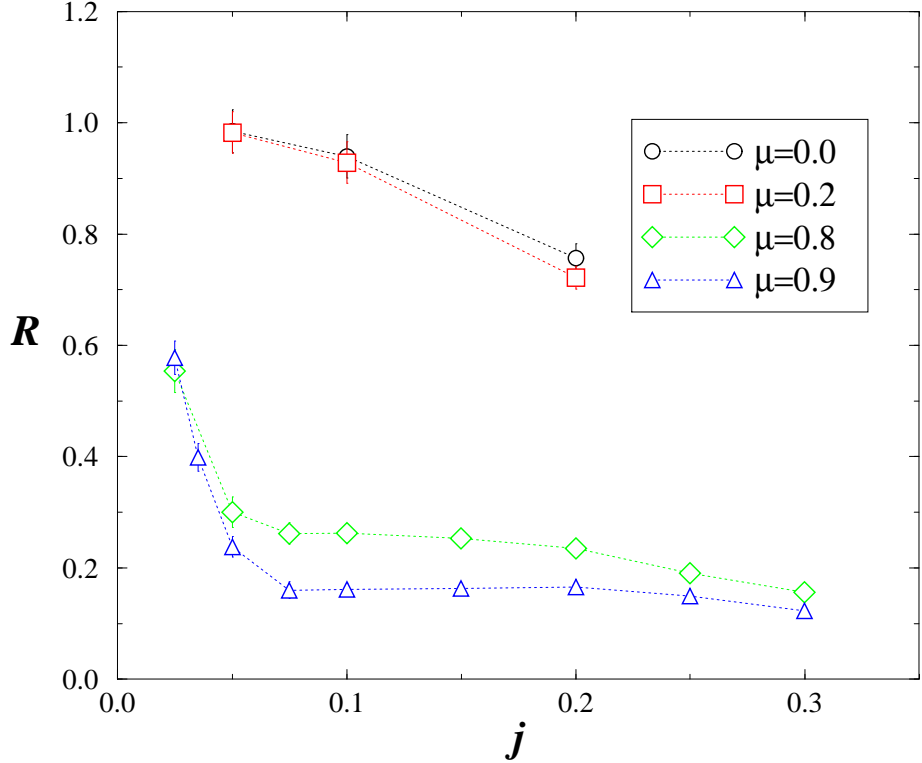, width =12cm}
\end{center}
\caption{
\label{fig:R}
Susceptibility ratios $R$ extrapolated to infinite volume for variuos $\mu=0.8$.
}
\end{figure}

The extrapolated results for $R$ as a function of $j$ are shown in
Fig.~\ref{fig:R}.
In the chirally broken phase the results support $R$ tending smoothly to one as
$j\to0$, consistent with unbroken baryon number symmetry. The behaviour in the
high-density phase is very different; the accumulation of intercepts in
Fig.~\ref{fig:R_extrap} manifests itself as a plateau for $j\gapprox0.075$.
For smaller $j$ the ratio shoots sharply upward towards one. This can
be attributed to a finite volume artifact, since we know that terms in
(\ref{eq:chil}) of the form $\langle qq(0)qq(x)\rangle$ which split
the degeneracy between $\chi_+$ and $\chi_-$ necessarily
vanish as $j\to0$ away from the thermodynamic limit. In this regard it is 
encouraging that these non-thermal effects manifest themselves in the
same range $j\lapprox0.05$ observed for the condensate measurements of
the previous section. We are thus motivated to attempt a linear extrapolation
to $j=0$ for the data with $j\in[0.75,0.2]$. The fits are of excellent
quality and yield 
$R(j=0)=0.29(2)$ for $\mu=0.8$ and $R(j=0)=0.17(1)$ for $\mu=0.9$.

Measurements of the diquark condensate 
of Sec.~\ref{subs:dqc} support a power-law form $\langle qq_+\rangle\propto
j^\alpha$ (\ref{eq:powerlaw}). If this is the case, then the relation
$\chi_+=\partial\langle qq_+\rangle/\partial j_+$ together with the Ward 
identity (\ref{eq:ward}) imply \cite{KKW}
\begin{equation}
R(j)={{\partial\ln\langle qq_+\rangle}\over{\partial\ln j_+}}=\alpha,
\end{equation}
which crucially is independent of $j$. The plateaux of Fig.~\ref{fig:R}
clearly support this interpretation; moreover the values we obtain for
$R(j=0)$ are in surprisingly good agreement with those from fits to
(\ref{eq:powerlaw}). The susceptibility measurements thus provide an independent
corroboration of the hypothesis that the system is critical for $\mu>\mu_c$.

\subsection{Diquark Masses}
\label{subs:diqm}

Our final numerical study in this sector is of the spatial behaviour of 
the diquark correlation functions, in an attempt to estimate the masses $M_\pm$
of diquark bound states. For brevity we only consider $\mu>\mu_c$, in which case
$M_\pm$ are probably best thought of as the energies required to excite 
a diquark pair above the ground state. We have restricted our attention to the
zero-momentum timeslice correlator 
$C_\pm(t)=\sum_{\vec x}\langle qq_\pm(\vec0,0)
qq_\pm(\vec x,t)\rangle$, so that the excited state 
must consist of quarks with equal
and opposite momentum $\vec k$. Recall that in the presence of a Fermi surface,
only quarks with $k\simeq k_F$ can be excited; the measurements presented here
are not sensitive to this restriction, although it will prove a decisive 
factor in the quasiparticle study
of Sec.~\ref{sec:quasi}. As in the previous section, we ignore ``singlet''
diagrams in calculating $C_\pm$.

\begin{figure}[p]
\begin{center}
\epsfig{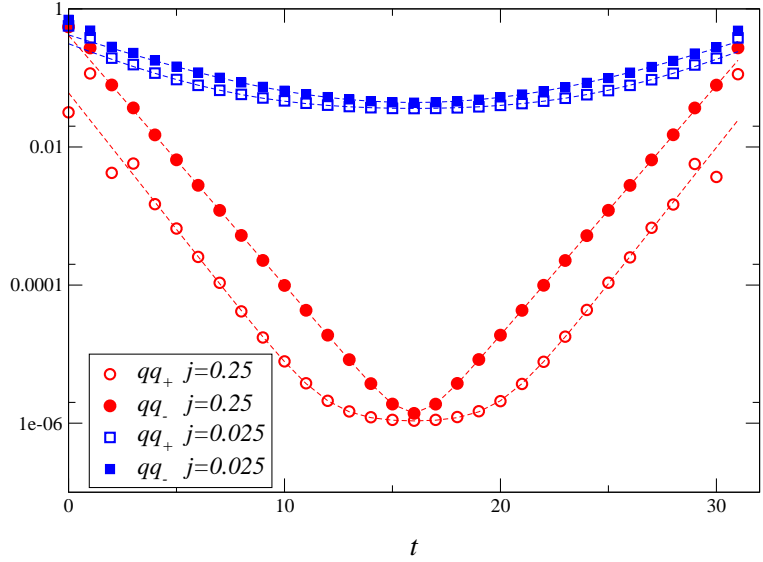}
\end{center}
\caption{
\label{fig:diquarks}
Diquark timeslice correlators $\vert C_\pm(t)\vert$ 
for two values of $j$ at $\mu=0.8$ on a
$32^3$ lattice.
}
\end{figure}
Fig.~\ref{fig:diquarks} shows the correlators for $j=0.025$ and 0.25 at
$\mu=0.8$. By virtue of its definition, $C_\pm$ is clearly symmetric
under time-reversal, in contrast to the correlators studied in \cite{HM}.
It is also clear, as expected, that $\vert C_-\vert>\vert
C_+\vert$, and that the difference between them grows with $j$.
Close inspection of Fig.~\ref{fig:diquarks} suggests that a standard simple-pole
fit to $C_\pm(t)$ will not succeed unless a constant term is included;
we have therefore attempted fits of the form \cite{HM}
\begin{equation}
C_\pm(t)=P_\pm\biggl(\exp(-M_\pm t)+\exp(-M_\pm(L_t-t))\biggr)+Q_\pm.
\label{eq:diqm}
\end{equation}
The plateau height $Q_+$ would by the cluster property be proportional to 
$\vert\langle qq_+\rangle\vert^2$ if the condensate formed; however,
the analysis of \cite{HM} showed that at $j=0$, $Q_+$ does not display
the required extensive scaling with two-dimensional spatial volume.
There is no obvious theoretical interpretation for $Q_-$.
\begin{figure}[p]
\begin{center}
\epsfig{file=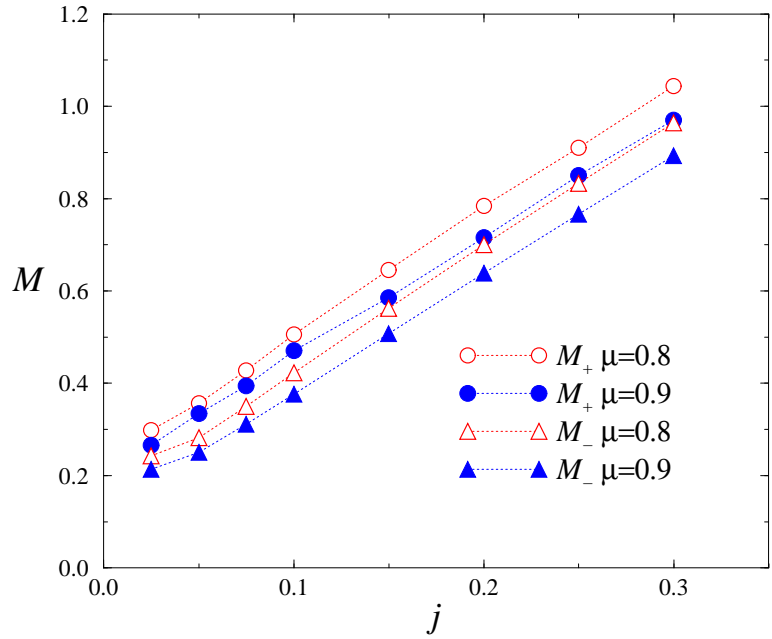, width =10cm}
\end{center}
\caption{
\label{fig:dqm}
Diquark masses extracted from fits to (\ref{eq:diqm}) on a $32^3$ lattice.
}
\end{figure}

Fig.~\ref{fig:dqm} shows $M_\pm(j)$ for $\mu=0.8,0.9$ resulting from fits of
(\ref{eq:diqm}) to timeslices 5 - 26. In most cases the $\chi^2/\mbox{dof}$
was $\lapprox2.0$ and in no case exceeded 6.0.
$M_\pm$ is found to
increase almost linearly with $j$, 
maintaining a roughly constant difference $M_+-M_-\simeq0.08$ for $j\geq0.05$.
For smaller $j$ the curvature in the plots suggests the two states may become
degenerate as $j\to0$.
A linear extrapolation to $j=0$ yields $M_+(\mu=0.8)\simeq0.23$, 
$M_+(\mu=0.9)\simeq0.21$, values of the same order of magnitude but
slightly lower than those obtained directly at $j=0$ on a $16^2\times24$ lattice
in \cite{HM} (note that the symbols $M_\pm$ have a different meaning in that
paper). The fitted values for $Q_\pm$ vary considerably over the range of $j$
explored: eg. for $\mu=0.8$, 
$Q_+$ rises from $0.102(1)\times10^{-5}$ at $j=0.25$
to $0.315(5)\times10^{-1}$ at $j=0.025$; in the same range $Q_-$ rises even more
dramatically from $0.49(6)\times10^{-7}$ to $0.279(7)\times10^{-1}$.

The most important feature of Fig.~\ref{fig:dqm} is that there is no evidence
for 
$M_-$ vanishing as $j\to0$, as might be expected if $qq_-$ coupled to a
Goldstone mode as a result of broken U(1)$_B$ symmetry. One might argue that 
the {\it ad hoc\/} 
inclusion of $Q_-$ in the fit (\ref{eq:diqm}) results in artificially high
values of $M_-$; in any case, the conclusion remains that simple pole fits to 
$C_-(t)$ corresponding to a weakly interacting
Goldstone boson in this channel fail drastically. The scaling form
(\ref{eq:powerlaw}) combined with the Ward identity ({\ref{eq:ward}) imples a
massless degree of freedom as $j\to0$, and hence long range correlations,
in both $qq_-$ and (since
symmetry is restored in this limit) $qq_+$ channels; they must, however, be 
strongly interacting and hence short-lived states.

\section{Criticality and Superfluidity}
\label{sec:crit}

Having established that in the limit of vanishing source 
there is no diquark condensation at high density,
but instead a critical phase with the scaling of the condensate with the source
governed by an
exponent $\delta$ varying continuously with $\mu$, 
we now discuss the implications
for possible superfluid behaviour of the 2+1$d$ NJL model. In fact, this result
is in accordance with well-known theorems that long range ordering of a 
two-dimensional system
with a continuous global symmetry is impossible \cite{MWC}. 
In the current context a particularly appropriate statement of the
theorem is due to Hohenberg \cite{Hohenberg}, who explicitly considers the case
of a composite order parameter via Cooper pairing in a low-dimensional fermion
superfluid.
Long-wavelength fluctuations of the phase $\theta$ of the would-be 
condensate always wash out the order in the zero source limit. 
In field theoretic language, in two
dimensions infra-red
divergences dictate that the Goldstone
pole in the transverse susceptibility 
predicted by a naive application of (\ref{eq:ward}) 
is replaced by a softer divergence
consistent with a power-law decay of the correlator
\begin{equation}
\lim_{j\to0}\langle
qq_-(0)qq_-(r)\rangle\propto
\langle e^{i\theta(0)}e^{-i\theta(r)}\rangle\propto {1\over r^\eta},  
\label{eq:spinwave}
\end{equation}
where $\eta$ is another critical exponent,
implying a massless but strongly-interacting mode and 
long-ranged phase correlations.
Note that direct numerical tests of
(\ref{eq:spinwave}) would 
require data from spatial diquark correlators, in contrast to the temporal
correlators explored in Sec.~\ref{subs:diqm}. There are also 
technical difficulties in taking the limit $j\to0$.

The best known example of a system with a critical phase is the 2$d$ O(2) spin
or $XY$ model, 
which is similar in that
long range order would also spontaneously break a U(1) global
symmetry. The critical behaviour occurs for $T<T_{BKT}$, the temperature of the
celebrated Berezinskii-Kosterlitz-Thouless transition \cite{KT,Ber}.
The physical picture can be explained as follows; on the assumption that 
the interaction strength is a periodic function
of the difference in angle $\theta$ between
adjacent spins, and is approximately gaussian in neighbourhood of its minima, 
then an effective Hamiltonian can be written as
\begin{equation}
H_{XY}[\theta,m]
=J\sum_{x\mu}(\partial_\mu\theta(x))^2-2\pi J\sum_{\tilde x,\tilde y}
m(\tilde x)\ln\left\vert{{\tilde x-\tilde y}\over r_0}\right\vert m(\tilde y),
\label{eq:XY}
\end{equation}
$J$ being the nearest neighbour coupling.
In addition to the $\theta$s the Hamiltonian depends on integer-charged
{\sl vortices\/} $m(\tilde x)$ located on the sites of the dual
lattice. The vortices are topological excitations of the spins which interact
via a Coulomb potential which is logarithmic in two dimensions, ensuring
that all configurations with finite $H_{XY}$ are overall charge-neutral, 
ie. contain as many anti-vortices as vortices.
The parameter $r_0$ is the ``core size'' of the vortex, which can be considered
of the same order as the lattice spacing.
Now, at low temperatures the second term of (\ref{eq:XY}) strongly suppresses
well-separated vortex -- anti-vortex pairs, and the model's dynamics are
dominated by small-amplitude fluctuations of $\theta$, the so-called {\sl spin
waves\/}. Phase correlations are governed by (\ref{eq:spinwave}) with 
$\eta(T)=T/4\pi J$, implying a critical phase with continuously varying
exponent. At the critical temperature $T_{BKT}$, the vortex entropy 
begins to dominate the free energy of the system, 
and vortex pairs of arbitrary separation form. The resulting
vortex plasma screens long range correlations resulting in a finite correlation
length for $T>T_{BKT}$.

Next we discuss the relation with superfluidity. We can rewrite the
diquark operator $qq_+(x)=\phi(x)=\phi_0e^{i\theta(x)}$, 
where the constant $\phi_0$ is
the density of quark pairs participating in the condensate and $\theta$ the 
local phase of the diquark operator. In this form $qq_+(x)$ can be regarded
as a bosonic {\sl macroscopic wavefunction\/} for the condensed pairs. 
We now identify
a superfluid current $J_{s\mu}$ via
\begin{equation}
J_{s\mu}\propto-{i\over2}\left(\phi^*\partial_\mu\phi-(\partial_\mu
\phi^*)\phi\right)
=K_s\partial_\mu\theta.
\label{eq:supercurrent}
\end{equation}
The constant $K_s$ must be determined empirically. In the non-relativistic
limit it is given by 
\begin{equation}
K_s={\hbar\over M}n_s
\end{equation}
where $M$ is the mass of the current-carrying atomic species
($M(^4\mbox{He})$ or $2M(^3\mbox{He})$ in the case of the two known 
superfluids), and $n_s$ a parameter called the {\sl superfluid density\/},
which for an interacting system 
need not coincide with the charge density of the particles in the condensate
\cite{Landau}. In turn this enables the definition of a {\sl superfluid
velocity\/} $\vec v_s={\hbar\over M}\vec\nabla\theta$. For a relativistic system
the relation $\vec v_s={\hbar\over2\mu}\vec\nabla\theta$ can
be shown to hold for diquark pairs for small $v_s$ \cite{IB}. 

Now in the static limit,
relation (\ref{eq:supercurrent}) implies that the flow is irrotational, 
viz. $\vec\nabla\times\vec J_s=\vec0$, and hence the circulation
$\kappa=\oint \vec J_s.\vec dl$ around any closed path vanishes unless either
the condensate is somewhere singular within the contour, ie. $\phi_0=0$,
or the space is non-simply connected. In either case the requirement that 
$\theta$ be single-valued results in the quantisation of circulation:
$\kappa=2\pi nK_s$, with $n$ integer. 
In the case of a singularity in $\phi$ the
physical realisation of $\kappa\not=0$ is a vortex, with a non-zero radius $r_0$
within which the normal phase is restored. Superfluid vortices experience
long-ranged mutual interactions; for a two-dimensional system such 
vortices can be identified with the vortices of the $XY$ model, and are expected
to be governed by an effective Hamiltonian of the same form as
(\ref{eq:XY}). An example of a non-simply connected space would be a finite
system of dimension $L_x\times L_y$ with periodic boundary conditions; 
in this case $\kappa\not=0$ implies a uniform supercurrent 
\begin{equation}
\vec J_s(n_x,n_y)=2\pi K_s
\left({n_x\over L_x}\hat{\vec x}+{n_y\over L_y}\hat{\vec y}\right).
\label{eq:uniform}
\end{equation}
The crucial point is that the resulting flow patterns are topologically stable,
implying the system's energy must be greatly increased 
to change $\kappa$ \cite{KT}.
For instance, in order to change $n_x$ by one unit, a 
vortex -- anti-vortex pair must be created and the vortex moved in the
$y$-direction right around the system 
before being allowed to reannihilate with the
anti-vortex. In so doing the system must be brought through a saddle-point
configuration in which the pair is separated by half the system extent;
from (\ref{eq:XY}) the energy required $\sim2\pi J\ln(L_y/2r_0)$. Since this 
diverges with the size of the system, we infer that the circulation is stable
and hence the current $\vec J_s$ persistent, implying
superfluidity. We conclude that the critical phase of the $XY$ model, and
by extension critical behaviour in any two dimensional system with a U(1)
global symmetry,
describes superfluidity despite the absence of a condensate. Long range order
of the phase $\theta$ is not necessary; phase coherence, as expressed by 
(\ref{eq:spinwave}), is sufficient. It is noteworthy in this regard that some
of the most precise tests of the universal predictions of the $XY$ model
have come from studies of thin films of superfluid $^4$He \cite{Nelson}.

Since we have used universal features of vortices and spin waves to 
argue that critical behaviour implies superfluidity in two dimensions, 
to justify the application of these ideas to the NJL model we should address
the issue of why our $\delta$ is not consistent with that of the
$XY$ model as revealed by a renormalisation group analysis \cite{Kosterlitz},
which predicts 
\begin{equation}
\delta\geq15\;\;\;;\;\;\;\eta\leq{1\over4},
\label{eq:XYdelta}
\end{equation}
with equality holding as $T\to T_{BKT-}$. First we note that dimensional
reduction, which predicts that the critical thermal
properties of 2+1$d$ systems should be governed by the $2d$
spin model with equivalent global symmetry, does {\sl not\/} apply in this case
(see \cite{Strouthos} for a recent numerical 
study of a four-fermi model with U(1) axial
symmetry at $T\not=0$ which does appear consistent with the BKT scenario).
Rather, the feature which permits us to use a two-dimensional effective model
is the static nature of the phase fluctuations, ie. $\partial_t\theta\simeq0$,
as evidenced by the plateaux observed in the large-$t$ behaviour
of diquark correlation functions in \cite{HM} and Sec.~\ref{subs:diqm}.
Note in addition that we have needed the limit $L_t\to\infty$ rather
than $L_t\to0$.
Since the number of accessible
Matsubara modes in our simulations remains large, 
the fermions need not decouple (indeed, there remain light fermionic
excitations, as we shall see in Sec.~\ref{sec:quasi}) and we
should not expect the model's dynamics to be 
described by a purely bosonic effective action.
Symmetry breaking via a composite order parameter is qualitatively different
from cases where the order parameter is an elementary field;
there is a wealth of evidence, both analytical and numerical,
that bosonic and fermionic models with the same patterns of global symmetry
breaking belong to separate universality classes in dimensions up to \cite{RWP,
HKK,Thirring, Jersak}
and even including \cite{HK} four.

Our simulations have yielded information only about the
critical exponent $\delta$ in the dense NJL model. Our results
$\delta(\mu=0.8)\approx3$, $\delta(\mu=0.9)\approx5$ are consistent with a 
critical phase for $\mu>\mu_c$. The lower numerical values as compared
to (\ref{eq:XYdelta}) typify the distinct nature of symmetry breaking
via a composite order parameter as discussed above.
Note, however, that although $\delta$ 
decreases with $\mu$, the analagous $\mu_{BKT}$ 
at which superfluid vortices unbind may not be physically accessible; most
probably chiral symmetry breaking at $\mu=\mu_c$ happens first.
Finally, although we have not yet found a method to measure $\eta$, it is 
interesting to estimate its value using the hyperscaling relation
$\delta=(d+2-\eta)/(d-2+\eta)$ \cite{Ma}. Since we have assumed
an effective dimension $d=2$ for the critical dynamics, the appropriate 
relation is
\begin{equation}
\delta={{4-\eta}\over\eta},
\end{equation}
yielding $\eta(\mu=0.8)\approx1$ and $\eta(\mu=0.9)\approx0.7$. Note that had
we used an effective dimension $d=3$, the prediction for $\eta$ at $\mu=0.9$ 
would be almost vanishing.

\section{The Quasiparticle Spectrum}
\label{sec:quasi}

In this section we study the spin-${1\over2}$ sector by examining the 
Gor'kov propagator ${\cal G}={\cal A}^{-1}$ as a function of $\mu$ and $j$. For
$\mu<\mu_c$ the fermion excitations are simply related to those at $\mu=0$, 
as reported in \cite{kim}. For $\mu>\mu_c$, however, the ground state of the
model changes radically, and is characterised by a Fermi surface with 
energy $E_F$ and characteristic momentum $k_F$. 
A generic description is Fermi liquid theory
\cite{Landau2,Landau}, in which excitations with momentum $k$ such that
$\vert k-k_F\vert\ll k_F$ 
are quasiparticles whose mass need not coincide with that of the 
fundamental quanta. If a BCS condensation occurs, then the lowest energy 
excitation may be separated from zero by a gap $\Delta$, and the quasiparticles
may not be eigenstates of baryon number, but instead some kind of particle-hole
superposition. Analysis of ${\cal G}$ using standard lattice spectroscopy
techniques yields information on the quasiparticle dispersion relation
$E(k)$, thus probing $\Delta$ and, more generally,
the nature of the model's Fermi surface.

We begin by making some general observations about the Gor'kov propagator.
Write
\begin{equation}
{\cal G}(x,y)=\left(\matrix{A(x,y)&N(x,y)\cr\bar N(x,y)&\bar A(x,y)\cr}\right),
\end{equation}
where each element denotes a $2\times2$ matrix in isospace. Our notation
signifies that the propagator contains both ``normal'' $\langle q(x)\bar
q(y)\rangle$ and ``anomalous'' $\langle q(x)q(y)\rangle$ components, together
with their barred counterparts. On a finite system, $A$ and $\bar A$ vanish in
the limit $j\to0$, and $N,\bar N$ become proportional to the usual 
fermion and anti-fermion
propagators. The number of independent components of ${\cal G}$ is
constrained by certain identities.
For instance, $N$ is proportional to an element of SU(2), implying
$N_{22}\equiv N_{11}^*$ and $N_{21}\equiv-N_{12}^*$, with similar relations for
$\bar N$. In the anomalous sector, however, it is $\tau_2A$ which 
resembles an SU(2) matrix so the equivalent identities 
are $A_{22}\equiv-A_{11}^*$ and $A_{21}\equiv A_{12}^*$. These relations
imply that a column of ${\cal G}$ can be reconstructed using just two conjugate
gradient inversions of ${\cal A}[\Phi]$.

We first examined the timeslice propagator 
${\cal G}(t)=\sum_{\vec x}{\cal G}(\vec0,0;\vec
x,t)$ 
and empirically found the following features:

\begin{itemize}

\item For $t\not=0$, 
$\mbox{Re}\langle N_{11}(t)\rangle\approx\mbox{Re}\langle\bar
N_{11}(L_t-t)\rangle$, ie. the antifermion propagator is related to
that of the fermion by time reversal.

\item 
$\mbox{Im}\langle N_{11}\rangle\approx
 \mbox{Im}\langle \bar N_{11}\rangle\approx
 \langle N_{12}\rangle\approx
 \langle \bar N_{12}\rangle\approx0$: the vanishing of the off-diagonal
components of $N$ is consistent with isopin SU(2)$_V$ symmetry.

\item $\mbox{Im}\langle A_{12}(t)\rangle\approx
       \mbox{Im}\langle \bar A_{12}(t)\rangle\approx
       -\mbox{Im}\langle A_{12}(L_t-t)\rangle$, ie. in the anomalous sector 
fermion and antifermion have equivalent behaviour under time-reversal.

\item $\mbox{Re}\langle A_{12}\rangle\approx\langle A_{11}\rangle\approx
                \langle\bar A_{11}\rangle\approx0$, ie. isopin symmetry
in the anomalous sector demands the diagonal components vanish. 

\end{itemize}

\begin{figure}[p]
\begin{center}
\vspace*{1.0cm}
\epsfig{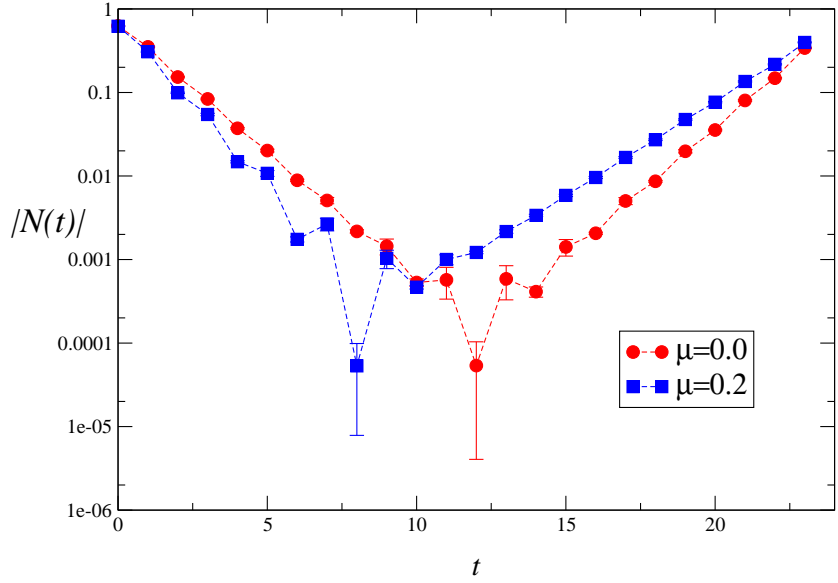}
\end{center}
\caption{
\label{fig:lowmu}
Normal propagators $\vert N(t)\vert$ for $\mu=0.0$ and $\mu=0.2$
on a $16^2\times24$ lattice with $j=0.1$.}
\end{figure}
From here on for convenience we will denote $\mbox{Re}N_{11}$ by $N$
and $\mbox{Im}A_{12}$ by $A$. 
In Fig.~\ref{fig:lowmu} we plot $\ln\vert N(t)\vert$ vs. $t$ for two values of 
$\mu$ in the low density chirally broken phase. At $\mu=0$ the propagator is
symmetric under time reversal, and the fermion mass $\Sigma\simeq0.730(2)$,
consistent with  
the breaking of chiral symmetry. At $\mu=0.2$ the time-reversal symmetry 
is broken; one state propagates forwards with mass 0.952(4), the other
backwards with mass 0.530(2), corresponding approximately 
to masses $\Sigma\pm\mu$. Now, since $n_B=0$ 
for $\mu<\mu_c$, the ground state is
unchanged and the physical interpretation of this result is simply that the 
chemical potential shifts the energies required to excite fermions and 
anti-fermions 
in opposite directions, the anti-fermion travelling in the $+t$
direction and the particle $-t$. 

For $\mu>\mu_c$ the situation is completely altered. Recall that in the presence
of a Fermi surface excitations have a characteristic momentum scale $k_F$.
Therefore it is necessary to introduce momentum dependence into the Gor'kov
propagator via ${\cal G}(\vec k,t)=\sum_{\vec x}
{\cal G}(\vec0,0;\vec x,t)e^{-i\vec
k.\vec x}$ \cite{HLM}. 
We choose $\vec k$ oriented along a lattice axis, and the set $\vec x$
to include only sites an even number of lattice spacings from the origin in each
direction, so that the physically accessible momenta are given by 
$k={2\pi n\over L_s}$ with $n=0,1,\ldots,L_s/4$.
Fig.~\ref{fig:highmu}
shows both normal and anomalous propagators in the high density phase for
$k={\pi\over4}$.
\begin{figure}[p]
\begin{center}
\epsfig{file=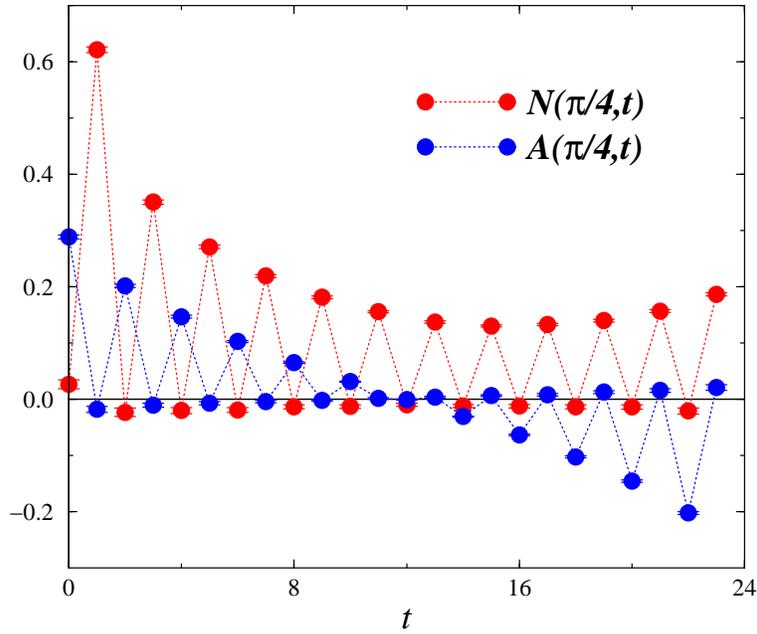, width =10cm}
\end{center}
\caption{
\label{fig:highmu}
Normal and anomalous propagators with $k={\pi\over4}$ at $\mu=0.8$
on a $16^2\times24$ lattice with $j=0.1$.}
\end{figure}
Note that now $N(t)\approx0$ for $t$ even, and $A(t)\approx0$ for $t$ odd.
This is a manifestation of the restored SU(2)$_L\otimes\mbox{SU(2)}_R$ symmetry
(\ref{eq:su2xsu2}), which would be broken by any $N_{ee,oo}$ or  
$A_{oe,eo}\not=0$. 
We have found that the propagators for various $j$ and $k$ are well fitted
on every second timeslice by the following forms, 
with fit parameters $A,B,C$ and $E$:
\begin{eqnarray}
N(k,t)&=&Ae^{-Et}+Be^{-E(L_t-t)},\label{eq:N(t)}\\
A(k,t)&=&C(e^{-Et}-e^{-E(L_t-t)})\label{eq:A(t)}.
\end{eqnarray}
The resulting $E(k,j)$ is the quasiparticle {\sl dispersion
relation\/}. We have studied this function in detail on a $32^3$ lattice,
which has 9 distinct values of $k$, performing fits to (\ref{eq:N(t)})
over the range $t\in[5,27]$. The fits are of excellent quality, with 
$\chi^2$/dof rarely exceeding 2.0.
We are naturally interested in the U(1)$_B$ symmetric limit; 
Fig.~\ref{fig:extrapE} shows that $E(k,j)$ can be smoothly 
extrapolated to $j=0$, and we quote the results of linearly extrapolating 
the data with $j\in[0.025,0.1]$.
\begin{figure}[htb]
\begin{center}
\epsfig{file=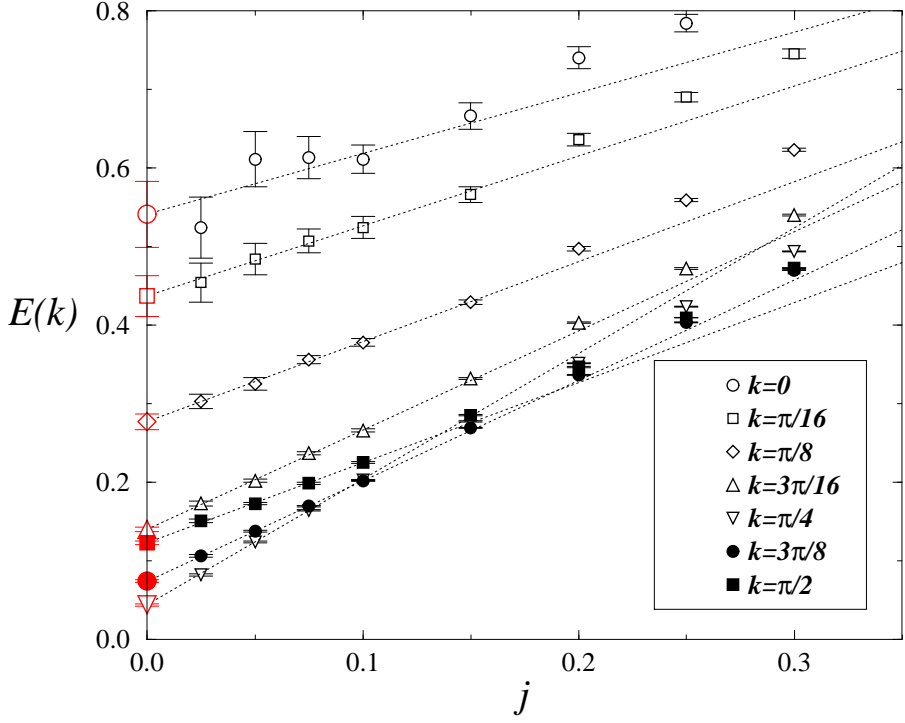, width =12cm}
\end{center}
\caption{
\label{fig:extrapE}
$E(k,j)$ vs. $j$ extracted from fits to (\ref{eq:N(t)})
on a $32^3$ lattice with $\mu=0.8$. Open symbols denote
excitations on the hole branch, and closed symbols the particle branch.}
\end{figure}
Fits to the anomalous propagator $A(k,t)$ yield quantitatively very 
similar results for $E(k,j)$.

The caption of Fig.~\ref{fig:extrapE} assigns different $k$ ranges to 
a ``hole branch'' ($E$ decreases with $k$) and a ``particle branch'' ($E$
increases with $k$).
It is straightforward to verify this interpretation by considering
the free Euclidean propagator $S_F(k)=(ik{\!\!\!/\,}+\mu\gamma_0+m)^{-1}$.
For fixed spatial momentum, with $\mu<E(k)=\sqrt{\vec k^2+m^2}$,
\begin{equation}
S_F(\vec k,t)=\cases{{m\over 2E}(1+V^-{\!\!\!\!\!\!\!\!/\;\;\;})
e^{-(E+\mu)t}&$t>0$;\cr
    {m\over 2E}(1-V^+{\!\!\!\!\!\!\!\!/\;\;\;})e^{-(E-\mu)\vert t\vert}
&$t<0$,\cr}
\label{eq:multE}
\end{equation}
where the complex ``4-velocity'' $V^\pm_\mu\equiv(E,\pm i\vec k)/m$. 
The propagator 
has both forwards and backwards-decaying signals, each associated with a 
different projection operator; in the limit $\vec k\to\vec0$ these become
${1\over2}(1\pm\gamma_0)$ 
and thus project onto anti-fermion and fermion 
states respectively. The fermion, being lighter, dominates
the signal for $\mu t\gg1$ yielding a predominantly backward propagation.
For $\mu>E(k)$, however, there is only a forwards-moving signal, once again 
dominated by the fermion:
\begin{equation}
S_F(\vec k,t)={m\over{2E}}\Theta(t)
\left[(1+V^-{\!\!\!\!\!\!\!\!/\;\;\;})e^{-(\mu+E)t} -
(1-V^+{\!\!\!\!\!\!\!\!/\;\;\;})e^{-(\mu-E)t}\right].
\label{eq:mugtE}
\end{equation}
For free fermions the transition between (\ref{eq:multE}) and (\ref{eq:mugtE}) 
takes place at a sharply defined Fermi energy $E_F=\sqrt{k_F^2+m^2}=\mu$.
Excitations with $k>k_F$, the Fermi momentum, 
add particles to levels above the Fermi surface; 
those with $k<k_F$
vacate holes in the Fermi Sea. The energy cost is smallest when $k\simeq
k_F$.

\begin{figure}[p]
\begin{center}
\epsfig{file=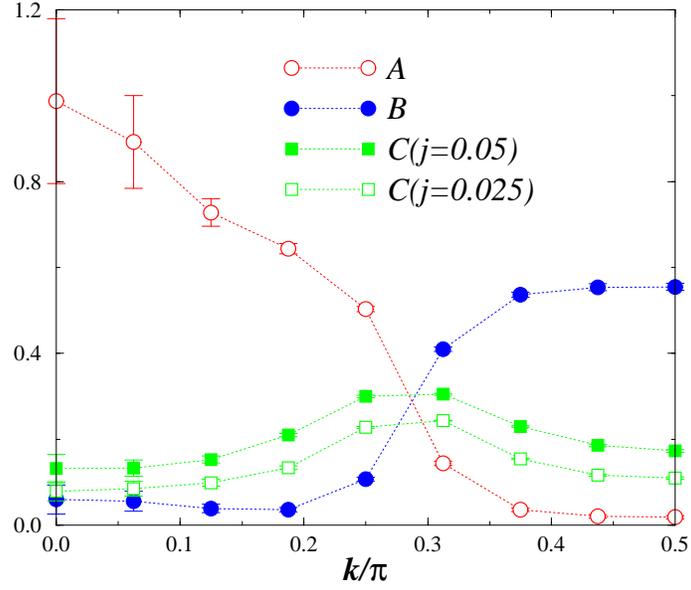, width =9cm}
\end{center}
\caption{
\label{fig:amp}
The amplitudes $A,B$ and $C$ extracted from fits to
(\ref{eq:N(t)},\ref{eq:A(t)}) 
on a $32^3$ lattice with $\mu=0.8$. The data for $A$ and $B$ were taken with
$j=0.025$.} 
\end{figure}
In Fig.~\ref{fig:amp} we plot the amplitudes $A,B$ and $C$ from the fits 
to (\ref{eq:N(t)},\ref{eq:A(t)}) and confirm that for small $k$, 
$N(k,t)$ is dominated by a
forwards-moving signal, but there is a rather sharp crossover to 
backwards propagation at $k/\pi\simeq0.3$. This transition becomes sharper
as $j\to0$; however, we plot data with $j\not=0$ to show that the 
amplitude $C$ only differs significantly from zero for momentum states in the
neighbourhood of the Fermi surface. Were a BCS gap to form, we would expect
$\lim_{j\to0}C(j)\not=0$ indicating particle-hole mixing; our data, however, 
do not give strong support for this.

\begin{figure}[p]
\begin{center}
\epsfig{file=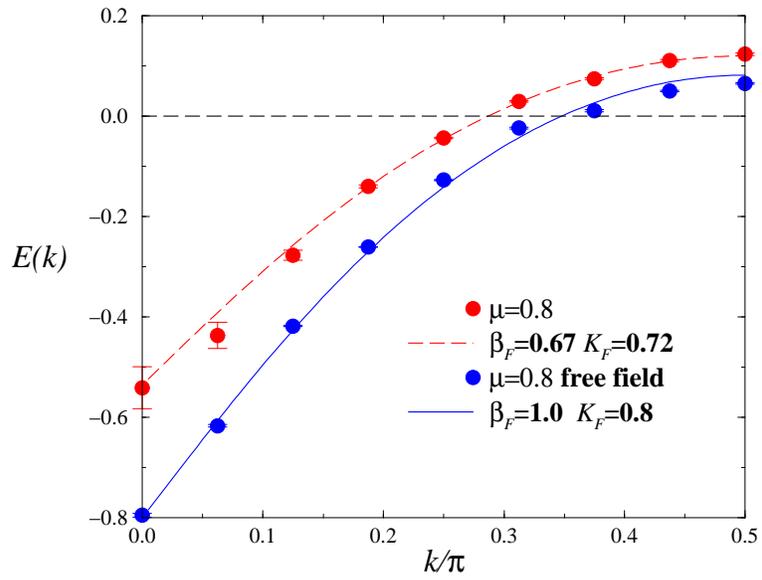, width =10cm}
\end{center}
\caption{
\label{fig:disperse}
Dispersion relation $E(k)$ at $\mu=0.8$ on a $32^3$ lattice for both 
interacting and free fermions.}
\end{figure}
Fig.~\ref{fig:disperse} shows the dispersion relation $E(k)$ extrapolated 
to $j=0$ for $\mu=0.8$, together with points derived from the free Gor'kov
propagator (ie. generated with $g^2=0$) using the identical procedure. We have
plotted energies from the hole branch as negative in order to generate a smooth
curve. There is no sign of any discontinuity characteristic of a BCS gap
$\Delta\not=0$. In order to interpret the detailed
form of the curve it is necessary to take account of discretisation effects;
for free massless fermions 
the expected dispersion relation, shown as a solid curve in
Fig.~\ref{fig:disperse} is $E(k)=-\mu+\sinh^{-1}(\sin k)$.
We have found that a reasonable fit to our data for $\mu\in[0.8,0.9]$,
shown as a dashed line in Fig.~\ref{fig:disperse},
is given by
\begin{equation}
E(k)=-E_0+D\sinh^{-1}(\sin k).
\label{eq:disperse}
\end{equation}
Eqn. (\ref{eq:disperse}) predicts a sharply defined effective Fermi momentum
given by 
\begin{equation}
K_F\equiv\sinh^{-1}(\sin k_F)=E_0/D. 
\end{equation}
In addition it is possible to define
a quasiparticle group velocity $\beta=\partial\sinh(E+E_0)/\partial\sin k$,
whose value 
\begin{equation}
\beta_F=D{{\cosh E_0}\over{\cosh K_F}}
\end{equation}
at the Fermi surface is the {\sl Fermi velocity\/}, 
which helps to characterise the Fermi liquid. For free massless fermions
$K_F=\mu$ and $\beta_F=1$ for all $\mu$. 
Our fitted values of $K_F$ and $\beta_F$ are given in
Table~\ref{tab:vf}.
\begin{table}[htb]
\caption{Quasiparticle parameters resulting from fits of (\ref{eq:disperse})
to data from a $32^3$ lattice. The quoted errors are purely statistical.} 
\label{tab:vf}
\begin{tabular*}{\textwidth}{@{}l@{\extracolsep{\fill}}lll}
\hline
$\mu$ &  $K_F$ & $\beta_F$ & $K_F/\mu\beta_F$ \\
\hline
0.80 & 0.720(3) & 0.670(3) &  1.34(1) \\
0.82 & 0.738(3) & 0.671(3) &  1.34(1) \\
0.84 & 0.773(3) & 0.686(3) &  1.34(1) \\
0.86 & 0.791(5) & 0.673(4) &  1.37(1) \\
0.88 & 0.811(5)& 0.628(4)  &  1.47(1) \\
0.90 & 0.836(4)& 0.704(4)  &  1.32(1) \\
\hline
\end{tabular*}
\end{table}

Although the errors quoted in Table~\ref{tab:vf} are almost certainly
underestimated, some systematic features are apparent. The observed values
of $K_F$ increase smoothly with $\mu$, and are $\approx90\%$ of their
free-field values. 
The Fermi velocity $\beta_F$, by contrast, is more or less
independent of $\mu$, and only $\approx70\%$ of the free-field value; ie. the 
quasiparticles travel at less than the speed of light.
In non-relativistic Fermi liquid theory \cite{Landau2,Landau}, the ratio 
$K_F/\beta_F$
defines a quantity called the {\sl effective mass\/} $M^*$, which need
not coincide with the mass of the fundamental atomic species $M$; eg. for the
archetypal Fermi liquid $^3$He in the sub-kelvin (but non-superfluid)
regime, $M^*\simeq3M$ \cite{Landau}. 
In Landau's theory the ratio for a two-dimensional
fluid is given in terms of the dipole component of the interaction between
quasiparticles, ie.
\begin{equation}
{M^*\over M}=1+N_F\int_0^{2\pi}\!{{d\vartheta}\over2\pi}\; f(\vartheta)\cos
\vartheta,
\label{eq:Landau}
\end{equation}
where $N_F=gV_2{k_F\over{2\pi\beta_F}}$ 
is the number of quasiparticle states on the Fermi surface
per unit energy interval ($V_2$ is the volume of $2d$ space and 
$g$ counts independent spin and isospin components) 
and $f(\vartheta)$ the spin-singlet 
interaction energy between 
quasiparticles at the Fermi surface with momenta separated by angle $\vartheta$.
In a relativistic generalisation the left hand side
of (\ref{eq:Landau}) is replaced by $K_F/\mu\beta_F$ \cite{Baym};
we infer from the data of Table~\ref{tab:vf} that 
$\overline{f(\vartheta)\cos\vartheta}>0$ when averaged over the Fermi circle,
and the interaction hence 
repulsive between quasiparticles with parallel momenta 
and/or attractive if the momenta are anti-parallel (the
simpler conclusion that the interaction is always attractive
was wrongly drawn in \cite{HLM}). 
This should be contrasted with the interaction between the fundamental
fermions of the NJL model due to $\sigma$ exchange, which  is attractive
and independent of direction\footnote
{Single $\pi$ exchange is attractive in isosinglet but repulsive in
isotriplet channels; the net binding effect in matter made from equal numbers of
$u$ and $d$ quarks vanishes.}. A similar effective reversal of sign arises
in the Hartree-Fock treatment of free electron states in a metal \cite{AM},
and is characteristic of a quantum-mechanical exchange effect.

To summarise, we have examined the quasiparticle spectrum and estimated both 
Fermi momentum $K_F$ and velocity $\beta_F$ with due allowance made for 
discretisation artifacts.
The results are consistent with a relativistic generalisation of a Landau Fermi
liquid, and qualitatively similar to the normal phase of liquid $^3$He.
There is no evidence for a BCS gap, and in the $j\to0$ limit the
anomalous components of the propagator signalling particle-hole mixing
probably vanish. We note, however, that
superfluid behaviour is not precluded by the absence of a gap \cite{Landau};
the unlimited growth of quasiparticle excitations that couple normal
and superfluid components and hence destroy
superfluidity in a gapless Bose liquid are here prevented by the Pauli
Exclusion Principle. Long range phase coherence is a sufficient condition for
superfluidity.

\section{Summary and Outlook}
\label{sec:conc}

Let us briefly review the main achievements of the paper. Firstly, we have
developed the necessary formalism to identify diquark condensation 
in numerical lattice studies of field theories on finite systems at
non-zero chemical potential, the crucial ingredient being the introduction of a
diquark source term. Secondly, to our initial surprise,
we have found no evidence for a condensate $\langle qq \rangle\not=0$ 
in studies of the 2+1$d$ NJL model in its high density phase $\mu>\mu_c$. 
Rather, the results from two independent analyses of diquark observables
are consistent with
a critical behaviour $\langle qq(j)\rangle\propto j^\alpha$
throughout the dense phase.
Whilst there is some residual uncertainty 
about the source of finite volume effects
for $j\lapprox0.05$, we suspect it would require computer resources considerably
greater than those we have used to modify this conclusion.
Critical behaviour in two dimensions implies long range coherence in the phase
of the condensate wavefunction, which is a sufficient condition for
superfluidity.
Whilst there is qualitative similarity with the well-known example of the low
temperature phase of the $2d$ $XY$ model, the measured value of
the critical exponent $\delta$ is distinct, suggesting that the 2+1$d$ dense 
NJL model belongs to a new universality class.
Thirdly, we have performed the first systematic spectroscopic study 
in the spin-${1\over2}$ sector for $\mu\not=0$, and mapped out the quasiparticle
dispersion relation. The success of the simple pole fits (\ref{eq:N(t)},
\ref{eq:A(t)})
confirms the long-lived nature of the quasiparticles. 
There is no evidence for either particle-hole mixing or
$\Delta\not=0$ in the $j\to0$ limit. Instead, the system resembles a normal
Fermi liquid with a well-defined Fermi surface; the Fermi velocity $\beta_F$ 
is of the same order as but 
significantly less than the
free-field value 1. Our findings can be summarised by the statement that the 
high density phase of the 2+1$d$ NJL model is a relativistic gapless thin film
BCS superfluid.

In a sense this and related papers with $\mu\not=0$ represent 
the primitive beginnings of the study of
condensed matter physics on the lattice. Let us sketch a few possible future
directions. Firstly, it would be interesting to estimate the 
supercurrent $\vec J_s$ corresponding to a quantised flow
pattern around a finite system as in (\ref{eq:uniform}), which could be set up
using a spatially varying $j(x)$. As well as providing a
direct demonstration of superfluidity, this would also enable the extraction of
the phenomenologically important parameter $K_s$. 
Secondly, it is possible to
study quasiparticles and other Fermi surface-related phenomena. 
For instance, a sharp 
Fermi surface leads to oscillations of spatial frequency $2k_F$ in 
the screened potential between static charges, known as Friedel oscillations
\cite{AM,Kapusta}. Friedel oscillations can be observed in the wavefunctions 
of $q\bar q$ and $qq$ states in 2+1$d$ four-fermi models with $\mu>\mu_c$
\cite{HKST}. Another possibility is the observation of light $q\bar q$
mesons in the 
spin-1 channel, corresponding to low-energy excitations of the Fermi surface
related to the phenomenon of zero sound \cite{Landau2,Landau,Langfeld}.
Finally, it is of prime importance to extend our calculations to the physically
relevant case of 3+1$d$. In this case the NJL model is no longer 
a fundamental field theory, but instead can be thought of as an effective
description of strong interaction physics with many possible phenomenological 
applications including thermodynamics \cite{Klevansky}.  If our arguments 
are correct, then the ``conventional'' signals for a BCS mechanism,
namely $\langle qq\rangle\not=0$ and $\Delta\not=0$ should be readily observed
using the methods we have developed.

\section*{Acknowledgements}

This work is supported
by the TMR network ``Finite temperature phase transitions in particle physics''
EU contract ERBFMRX-CT97-0122. SJH also received support from the Leverhulme
Trust.


\begin{thebibliography}{xx}
%
\bibitem{BCS} J. Bardeen, L.N. Cooper and J.R. Schrieffer, Phys. Rev. {\bf108}
(1957) 1175.
%
\bibitem{Gorkov} L.P. Gor'kov, Sov. Phys. JETP {\bf7} (1958) 505.
%
\bibitem{NJL} Y. Nambu and G. Jona-Lasinio, Phys. Rev. {\bf122} (1961) 345;
{\it ibid\/} {\bf124} (1961) 246.
%
\bibitem{diquarks} B.C. Barrois, Nucl. Phys. {\bf B129} (1977) 390;\\
D. Bailin and A. Love, Phys. Rep. {\bf107} (1984) 325;\\
M. Alford, K. Rajagopal and F. Wilczek, Phys. Lett. {\bf B422} (1998) 247;
Nucl. Phys. {\bf B537} (1999) 443;\\
R. Rapp, T. Sch\"afer, E.V. Shuryak and M. Velkovsky, Phys. Rev. Lett. {\bf81}
(1998) 53.
%
\bibitem{Berges} J. Berges and K. Rajagopal, Nucl. Phys. {\bf B538} (1999) 215.
%
\bibitem{Pis} R.D. Pisarski and D.H. Rischke, Phys. Rev. {\bf D60}:094013
(1999); {\it ibid\/} {\bf D61}:051501 (2000); {\bf D61}:074017
(2000).
%
\bibitem{KSTVZ} J.B. Kogut, M.A. Stephanov, D. Toublan, J.J.M. Verbaarschot and
A. Zhitnitsky, Nucl. Phys. {\bf B582} (2000) 477.
%
\bibitem{istvan}
S.J. Hands, I. Montvay, S.E. Morrison, M. Oevers, L. Scorzato and J. Skullerud,
Eur. Phys. J. {\bf C17} (2000) 285.
%
\bibitem{KSHM} J.B. Kogut, D.K. Sinclair, S.J. Hands and S.E. Morrison, 
{\tt hep-lat/0105026}.
%
\bibitem{Barbour} I.M. Barbour, S.J. Hands, J.B. Kogut, M.-P. Lombardo and
S.E. Morrison, Nucl. Phys. {\bf B557} (1999) 327.
%
\bibitem{kim}
S.J. Hands, S. Kim and J.B. Kogut, Nucl. Phys. {\bf B442} (1995) 364.
%
\bibitem{TT} D.R. Tilley and J. Tilley, {\sl Superfluidity and
Superconductivity\/}, (Adam Hilger, Bristol, 1990).
%
\bibitem{AM}
N.W. Ashcroft and N.D. Mermin, {\sl Solid State Physics\/}, (Holt, Rinehart and
Winston, New York, 1976).
%
\bibitem{Volovik} G.E. Volovik, {\sl Exotic Properties of Superfluid $^3$He\/}
(World Scientific, Singapore, 1992).
%
\bibitem{HM} S.J. Hands and S.E. Morrison, Phys. Rev. {\bf D59}:116002 (1999).
%
\bibitem{HLM} S.J. Hands, B. Lucini and S.E. Morrison, Phys. Rev. Lett.
{\bf86} (2001) 753.
%
\bibitem{RWP} B. Rosenstein, B.J. Warr and S.H. Park, Phys. Rep. {\bf205} (1991)
59.
%
\bibitem{HKK} S.J. Hands, A. Koci\'c and J.B. Kogut, Ann. Phys. (N.Y.) {\bf224}
(1993) 29.
%
\bibitem{Nelson} D.R. Nelson and J.M. Kosterlitz, Phys. Rev. Lett. {\bf 39}
(1977) 1201;\\
D.R. Nelson, in {\sl Phase Transitions and Critical Phenomena\/}, Vol {\bf7}
(1983) p.1, eds. C. Domb and J.L. Lebowitz (Academic Press, London).
%
\bibitem{KT} J.M. Kosterlitz and D.J. Thouless, J. Phys. {\bf C6} (1973) 1181. 
%
\bibitem{Landau2}
L.D. Landau, Sov. Phys. JETP {\bf3} (1956) 920; {\it ibid\/} {\bf5} (1957) 101.
%
\bibitem{Landau} E.M. Lifshitz and L.P. Pitaevskii, {\sl Statistical Physics
(Part 2)\/} (Landau and Lifshitz Vol. 9) (Pergamon Press, Oxford 1980).
%
\bibitem{HK} S.J. Hands and J.B. Kogut, Nucl. Phys. {\bf B520} (1998) 382.
%
\bibitem{RWP2} B. Rosenstein, B.J. Warr and S.H. Park, Phys. Rev. {\bf D39}
(1989) 3088;\\
S.J. Hands, A. Koci\'c and J.B. Kogut, Nucl. Phys. {\bf B390} (1993) 355;\\
A.S. Vshivtsev, B.V. Magnitskii, V.Ch. Zhukovskii and K.G. Klimenko, 
Phys. Part. Nucl. {\bf29} (1998) 523.
%
\bibitem{Costas} J.B. Kogut and C.G. Strouthos, Phys. Rev. {\bf D63}:054502 
(2001).
%
\bibitem{BB}
C.J. Burden and A.N. Burkitt, Europhys. Lett. {\bf3} (1987) 545.
%
\bibitem{ray}
S. Gottlieb, W. Liu, D. Toussaint, R.L. Renken and R. Sugar, Phys. Rev. {\bf
D35} (1987) 2531.
%
\bibitem{Ma} S.-K. Ma, {\sl Modern Theory of Critical Phenomena\/}, (Benjamin,
Reading MA, 1976).
%
\bibitem{Thirring} L. Del Debbio, S.J. Hands and J.C. Mehegan, Nucl. Phys. {\bf
B502} (1997) 269.
%
\bibitem{KKW} A. Koci\'c, J.B. Kogut and K.C. Wang, Nucl. Phys. {\bf B398}
(1993) 405.
%
\bibitem{MWC} N.D. Mermin and H. Wagner, Phys. Rev. Lett. {\bf17} (1966) 1133;\\
S. Coleman, Comm. Math. Phys. {\bf31} (1973) 259.
%
\bibitem{Hohenberg} P.C. Hohenberg, Phys. Rev. {\bf158} (1967) 383.
%
\bibitem{Ber} V.L. Berezinskii, Sov. Phys. JETP {\bf34} (1972) 610.
%
\bibitem{IB} K. Iida and G. Baym, {\tt hep-ph/0108149}.
%
\bibitem{Kosterlitz} J.M. Kosterlitz, J. Phys. {\bf C7} (1974) 1046.
%
\bibitem{Strouthos} S.J. Hands, J.B. Kogut and C.G. Strouthos, Phys. Lett.
{\bf B515} (2001) 407. 
%
\bibitem{Jersak} J.A. Gracey, Int. J. Mod. Phys. {\bf A6} (1991) 395; {\it
ibid\/} {\bf A9} (1994) 567, Phys. Rev. {\bf D50} (1994) 2840;\\
L. K\"arkk\"ainen, R. Lacaze, P. Lacock and B. Petersson, Nucl. Phys. {\bf B415}
(1994) 781;\\
E. Focht, J. Jers\'ak and J. Paul, Phys. Rev. {\bf D53} (1996) 4616;\\
S.J. Hands and B. Lucini, Phys. Lett. {\bf B461} (1999) 263.
%
\bibitem{Baym} G. Baym and S.A. Chin, Nucl. Phys. {\bf A262} (1976) 527.
%
\bibitem{Kapusta}
J. Kapusta and T. Toimela, Phys. Rev. {\bf D37} (1988) 3731.
%
\bibitem{HKST}
S.J. Hands, J.B. Kogut, C.G. Strouthos and T.N. Tran, in preparation.
%
\bibitem{Langfeld} K. Langfeld, H. Reinhardt and M. Rho, Nucl. Phys. {\bf A622}
(1997) 620;\\
K. Langfeld, Nucl. Phys. {\bf A642} (1998) 96c.
%
\bibitem{Klevansky}
S.P. Klevansky, Rev. Mod. Phys. {\bf64} (1992) 649.



\end{thebibliography}
\end{document}